\documentclass[fleqn,usenatbib]{mnras}

\usepackage{newtxtext,newtxmath}
\usepackage{threeparttable}
\usepackage[T1]{fontenc}
\usepackage{ae,aecompl}
\DeclareRobustCommand{\VAN}[3]{#2}
\let\VANthebibliography\thebibliography
\def\thebibliography{\DeclareRobustCommand{\VAN}[3]{##3}\VANthebibliography}

\usepackage{graphicx}	
\usepackage{amsmath}	
\usepackage{soul}	
\usepackage{color,xcolor}	
\usepackage{float}
\usepackage{subfig}
\usepackage{pdflscape}
\usepackage{soul} 

\title[Physical properties of $Fermi$ LBLs]{The physical properties of $Fermi$-4LAC low-synchrotron-peaked BL Lac objects}

\author[H.-B. Hu et al.]{
Hai-Bin Hu,$^{1,4}$
Hai-Qin Wang,$^{2,4}$
Rui Xue,$^{1}$\thanks{E-mail: ruixue@zjnu.edu.cn}
Fang-Kun Peng,$^{2}$\thanks{E-mail: pengfk@ahnu.edu.cn}
and Ze-Rui Wang$^{3}$\thanks{E-mail: zerui\_wang62@163.com}
\\
$^{1}$Department of Physics, Zhejiang Normal University, Jinhua 321004, People’s Republic of China\\
$^{2}$Department of Physics, Anhui Normal University, Wuhu 241002, People’s Republic of China\\
$^{3}$College of Physics and Electronic Engineering, Qilu Normal University, Jinan 250200, People’s Republic of China\\
$^{4}$These authors contributed equally to this work.
}

\date{Accepted XXX. Received YYY; in original form ZZZ}

\pubyear{2023}

\begin{document}
\label{firstpage}
\pagerange{\pageref{firstpage}--\pageref{lastpage}}
\maketitle

\begin{abstract}
Previous studies on the fitting of spectral energy distributions (SEDs) often apply the external-Compton process to interpret the high-energy peak of low-synchrotron-peaked (LSP) BL Lac objects (LBLs), despite the lack of strong broad emission lines observed for LBLs. In this work, we collect quasi-simultaneous multi-wavelength data of 15 LBLs from the $Fermi$ fourth LAT AGN catalog (4LAC). We propose an analytical method to assess the necessity of external photon fields in the framework of one-zone scenario. Following derived analytical results, we fit the SEDs of these LBLs with the conventional one-zone leptonic model and study their jet physical properties. Our main results can be summarized as follows. (1) We find that most LBLs cannot be fitted by the one-zone synchrotron self-Compton (SSC) model. This indicates that external photons play a crucial role in the high-energy emission of LBLs, therefore we suggest that LBLs are masquerading BL Lacs. (2) We suggest that the $\gamma$-ray emitting regions of LBLs are located outside the broad-line region and within the dusty torus. (3) By extending the analytical method to all types of LSPs in $Fermi$-4LAC (using historical data), we find that the high-energy peaks of some flat spectrum radio quasars and blazar candidates of unknown types can be attributed to the SSC emission, implying that the importance of external photons could be minor. We suggest that the variability timescale may help distinguish the origin of the high-energy peak.
\end{abstract}




\section{Introduction}

Blazars are the most extreme subclass of active galactic nuclei (AGNs),
whose emission is dominated by the non-thermal emission
from the relativistic jet orients close to observers' line of sight
\citep{1995PASP..107..803U}. Due to relativistic beaming effects,
the observed emission is Doppler boosted,
and variability timescale is shortened \citep[e.g.,][]{2019Galax...7...20B, 2020Galax...8...72C}.
The spectral energy distributions (SEDs) of blazars usually show
a double-peak structure in the $\rm log\nu - \rm log\nu\it F_{\nu}$
diagram \citep[e.g.,][]{2007Ap&SS.309...95B}.
Because of the detected significant linear polarization, the low-energy peak
(from radio to UV/X-ray) is attributed to synchrotron
emission from relativistic electrons in a magnetic field.
In leptonic scenarios, the high-energy peak
(from X-ray to $\rm\gamma$-ray band) is
attributed to the inverse Compton (IC) scattering of relativistic electrons
\citep[e.g.,][]{1995ApJ...446L..63D, 2002ApJ...575..667D, 2007Ap&SS.309...95B}.
The seed photons for the IC process originate from the synchrotron
emission of the same population of relativistic electrons
\citep[synchrotron self-Compton, SSC; e.g.,][]{1998ApJ...509..608T}
or from external photon fields
\citep[external-Compton, EC; e.g.,][]{2009MNRAS.397..985G},
such as the accretion disk \citep[e.g.,][]{1993ApJ...416..458D},
the broad-line region
\citep[BLR,][]{1994ApJ...421..153S, 2006ApJ...646....8F}
and the dusty torus
\citep[DT,][]{2000ApJ...545..107B, 2005ApJ...629...52S}.
Alternatively, the hadronic model is another possible
origin of the high-energy peak 
\citep[e.g.,][]{1993A&A...269...67M, 2012ApJ...755..147D, 2012A&A...546A.120D,
2022A&A...659A.184L, 2022PhRvD.106j3021X, 2023PhRvD.107j3019X,
2023arXiv230810200W}.
Based on the emission line features, blazars can be classified into BL Lac
objects (BL Lacs) with no or weak emission lines (equivalent width, $EW < 5$~\AA)
and flat spectrum radio quasars (FSRQs) with strong emission lines 
\citep[$EW\geqslant5$~\AA;][]{1995PASP..107..803U}.
According to the peak frequency of the low-energy synchrotron peak ($\nu_{\rm s}$),
\citet{2010ApJ...716...30A} divided blazars into low-synchrotron-peaked
(LSP, $\nu_{\rm s}\lesssim10^{14}$ Hz), intermediate-synchrotron-peaked (ISP, $10^{14}\lesssim\nu_{\rm s}\lesssim10^{15}$ Hz)
and high-synchrotron-peaked (HSP, $\nu_{\rm s}\gtrsim10^{15}$ Hz) sources.
\citet{2022ApJS..263...24A} found
that FSRQs are basically LSPs, while BL Lacs have a more uniform distribution of
low-energy peaks. Modeling SEDs of blazars provides us a way to explore
the intrinsic physical properties of emitting region inside the jet
\citep[e.g.,][]{2008MNRAS.387.1669G, 2009MNRAS.397..985G, 2010MNRAS.409L..79G, 2014Natur.515..376G, 2011ApJ...733...14M, 2012ApJ...753..154M, 2020ApJS..248...27T, 2021MNRAS.506.5764D, 2014MNRAS.439.2933Y, 2012ApJ...752..157Z}.

The conventional one-zone leptonic model has been widely applied to reproduce
blazars' SEDs. FSRQs are usually fitted with a one-zone EC model in their SEDs 
\citep[e.g.,][]{2010MNRAS.402..497G, 2014Natur.515..376G}, due to the observation of broad
emission lines. In contrast, the situation for BL Lacs is more complicated.
ISP BL Lacs (IBLs) and HSP BL Lacs (HBLs) generally adopt a one-zone SSC model
for fitting \citep[e.g.,][]{2012ApJ...752..157Z, 2014MNRAS.439.2933Y, 2017MNRAS.464..599D},
while LSP BL Lacs (LBLs) typically use the one-zone EC model
\citep[e.g.,][]{2000AJ....119..469B, 2021RAA....21..305W, 2023MNRAS.521.6210D}. \citet{2014MNRAS.439.2933Y} found that the one-zone SSC model has difficulty in
explaining the SEDs of LBLs. Specifically, the model predicted low-energy peaks
are higher than those suggested by the data points.
Furthermore, the X-ray spectrum of some LBLs is too soft to naturally extend to
the higher-energy band. Therefore it is reasonable to introduce another component
to interpret the high-energy peak. In the one-zone leptonic model,
an EC component is typically introduced, even though
no strong broad emission lines are observed for LBLs.
If the EC emission does indeed dominate the high-energy peak,
then it implies that the broad emission lines of LBLs are outshone by the luminous
non-thermal emission from the jet.
Consequently, LBLs can be considered as masquerading BL Lacs
\citep[e.g.,][]{2012MNRAS.420.2899G, 2013MNRAS.431.1914G, 2019PhRvD..99f3008L, 2022ApJ...936..146X}.
Therefore, it would be advantageous to explore if external photons are
indispensable for interpreting the high-energy peak of LBLs.

The rapid and large-amplitude variability of blazars
imply that not only the flux in each band vary dramatically,
but also the peak frequencies of the double peaks shift significantly with time 
\citep[e.g.,][]{2004A&A...413..489M, 2021MNRAS.502..836S, 2012ApJ...752..157Z}.
Therefore, quasi-simultaneous multi-wavelength SEDs are
essential to reveal the physical properties of their jets
\citep[e.g.,][]{2014MNRAS.439.2933Y}.
In this work, we collect quasi-simultaneous multi-wavelength data
of 15 LBLs from the $Fermi$ fourth LAT AGN catalog (4LAC).
By using an analytical method to constrain the parameter space,
we analyse the applicability of the one-zone SSC model to LBLs.
We explore the physical properties of LSP blazars jets by interpreting
their SEDs in the framework of one-zone leptonic model.
This paper is organized as follows.
In Section~\ref{sample}, we present the sample collection and $Fermi$-LAT data analysis.
The analytical method and model description are given in Section~\ref{Analytical and model},
followed by the results and discussion in Section~\ref{result}.
Finally, we end with conclusions in Section~\ref{conclusion}.
The cosmological parameters $H_{0}=69.6\ \rm km\ s^{-1}Mpc^{-1}$,
$\Omega_{0}=0.29$, and $\Omega_{\Lambda}$= 0.71 
\citep{2014ApJ...794..135B} are used throughout this work.

\section{sample collection and data analysis}\label{sample}
\subsection{Sample collection}
We collected 15 LBLs from $Fermi$-4LAC as our sample. To obtain their contemporaneous multi-wavelength SEDs, we firstly collected all data and corresponding observation times of these LBLs
in the Space Science Data Center (SSDC) SED Builder
\citep{2011arXiv1103.0749S}\footnote{\url{http://tools.ssdc.asi.it/SED/}}.
Then, we searched for the intersection of observation time
in the infrared to X-ray range and ensured that the time interval between
any two bands does not exceed seven days.
The radio data were not taken into account, because they could not be explained
by a one-zone model and the radio data available
on the SSDC were mostly collected around the 1990s.
Finally, we obtained the $Fermi$ GeV data by integrating two months that include the previous time intersection (details can be found in Section~\ref{fermi}), because the $\rm\gamma$-ray data
on the SSDC are obtained by integrating annually, which obviously does not meet the requirement of contemporaneous multi-wavelength data. Note that the observation time intervals of the data from
infrared to X-ray range that we collected are within one week,
while the $Fermi$-LAT data are integrated for two months.
Therefore, the multi-wavelength data in this sample are quasi-simultaneous
rather than simultaneous. The details of the quasi-simultaneous data
of each LBL are given in Table~\ref{table:sample}.

\subsection{\textsl{Fermi}-LAT data analysis}\label{fermi}
The LAT on board the \textsl{Fermi} mission is a pair-conversion instrument that is sensitive to GeV emission \citep{2009ApJ...697.1071A}. We collected \textsl{Fermi}-LAT data in the sky-survey mode encompassing two months listed as $t_2$ in Table~\ref{table:sample}, which is taken $t_1$ from the Table~\ref{table:sample} as the midpoint. Data were analyzed with the fermitools version 2.2.0. A binned maximum likelihood analysis was performed on a region of interest (ROI) with a radius $10^{\circ}$ centered on the ``R.A.” and “decl.'' of each source. Recommended event selections for data analysis were ``FRONT+BACK'' ({evtype=3}) and evclass=128. We applied a maximum zenith angle cut of $z_{\rm zmax} = 90^{\circ}$ to reduce the effect of the Earth albedo background. The standard gtmktime ﬁlter selection with an expression of (DATA\_QUAL $> 0\ \&\& \ $ LAT\_{CONFIG} $== 1$) was set. A source model was generated containing the position and spectral deﬁnition for all the point sources and diffuse emission from the 4FGL \citep{2022ApJS..260...53A} within $15^{\circ}$ of the ROI center. The analysis included the standard galactic diffuse emission model ($\rm gll\_iem\_v07.ﬁts$) and the isotropic component (iso\_P8R3\_SOURCE\_V3\_v1.txt), respectively. We binned the data in counts maps with a scale of $0.1^{\circ}$ per pixel and used 30 logarithmically spaced bins in energy of $0.1 - 100$ GeV. The energy dispersion correction was made when event energies extending down to 100 MeV were taken into consideration. We divide this spectral energy distribution into six or three equal logarithmic energy bins in the $0.1 - 100$ GeV for sources above or below $TS = 25$, respectively, shown in Fig.~\ref{fig2}. The data points with $TS < 4$ or nominal flux uncertainty larger than half the flux itself are given upper limits at the 95\% conﬁdence level.

\begin{table*}
\scriptsize 
\caption{Details of the quasi-simultaneous data of our sample. Columns from left to right: (1) source name in the $Fermi$ catalog; (2) source name; (3) redshift; (4) right ascension (RA); (5) declination (Dec.); (6) observation time from infrared to X-ray bands; (7) integration time of $\gamma$-rays.}\label{table:sample}
\centering
\begin{tabular}{ccccccc}
\hline\hline
$Fermi$ name & Source name & $z$ & RA (J2000) & Dec. (J2000) & $t_{1}$ & $t_{2}$\\
& & & (degrees) & (degrees) & &\\
~(1) & (2) & (3) & (4) & (5) & (6) & (7)\\
\hline

4FGL J0100.3+0745   & GB6 J0100+0745 & 0.30052  &     15.0866     &     7.7643      &      2010.07.09       & 2010.06.09--2010.08.09\\
4FGL J0141.4-0928   &   PKS 0139-09  &  0.733   &    25.357634    &    -9.478798    & 2010.05.30--2010.06.06 & 2010.05.03--2010.07.03\\
4FGL J0210.7-5101   &  PKS 0208-512  &  1.003   &    32.692502    &   -51.017193	&      2009.11.26       & 2009.10.26--2009.12.26\\
4FGL J0238.6+1637   &  PKS 0235+164  &  0.94    &    39.662209    &    16.616465	&      2010.01.30       & 2010.01.01--2010.03.01\\
4FGL J0334.2-4008   &  PKS 0332-403  &  1.445   &    53.556894    &   -40.140388	& 2010.01.17--2010.01.18 & 2009.12.17--2010.02.17\\
4FGL J0522.9-3628   &   PKS 0521-36  &  0.055   &    80.741603    &    -36.45857	&      2010.03.05       & 2010.02.05--2010.04.05\\
4FGL J0854.8+2006   &  PKS 0851+202  &  0.306   &    133.703646   &    20.108511	&      2010.04.10       & 2010.03.10--2010.05.10\\
4FGL J0958.7+6534   &   S4 0954+65   &  0.367   &   149.696855    &    65.565228	&      2010.03.12       & 2010.02.12--2010.04.12\\
4FGL J1043.2+2408   &   B2 1040+24A  & 0.559117 &   160.787649    &    24.143169	&      2010.07.09       & 2010.06.09--2010.08.09\\
4FGL J1147.0-3812   &  PKS 1144-379  &  1.048   &   176.755711    &   -38.203062	&      2010.06.24       & 2010.05.24--2010.07.24\\
4FGL J1517.7-2422   &     AP Lib     &  0.048   &    229.424223   &   -24.372078	&      2010.02.20       & 2010.01.20--2010.03.20\\
4FGL J1751.5+0938   &     OT 081     &  0.322   &   267.886744    &    9.650202 	&      2010.04.01       & 2010.03.01--2010.05.01\\
4FGL J1800.6+7828   &  S5 1803+784   &  0.68    &    270.190349   &    78.467783	&      2009.10.13       & 2009.09.13--2009.11.13\\
4FGL J2152.5+1737   &   S3 2150+17   &  0.871   &     328.137     &     17.6173	    & 2010.04.08--2010.04.10 & 2010.03.09--2010.05.09\\
4FGL J2247.4-0001   &  PKS 2244-002  &  0.949   &     341.867     &     -0.0263     & 2010.01.14--2010.01.16 & 2009.12.15--2010.02.15\\

\hline
\end{tabular}
\end{table*}

\section{Analytical method and model description}\label{Analytical and model}

The following analytical method and numerical modeling are
carried out within the framework of the
one-zone leptonic model. It is assumed that all the radiation of the blazar
jet comes from a spherical emitting region with radius $R$, which is filled with a
uniform magnetic field $B$ and a plasma of charged particles. The emitting region
moves along a relativistic jet with the bulk Lorentz factor
$\Gamma=(1-\beta^2)^{-1/2}$
at a viewing angle $\theta \mathrm{^{obs} }$ to the
observer's line of sight, where $\beta c$
is the speed of the emitting region.
Due to relativistic beaming effects, the observed flux is boosted
by a factor of $\delta ^{4} $, where
$\delta =\left [ \Gamma \left ( 1-\beta \cos \theta ^{\mathrm{obs}  }  \right )  \right ] ^{-1}$
is the Doppler factor. In this paper, we approximate $\delta \approx \Gamma$
by assuming $\theta \mathrm{^{obs} }  \lesssim1 /\Gamma$. In this section, parameters without superscript are measured in the comoving frame,
those with superscript "obs" are measured in the frame of the observer, and those
with superscript "AGN" are measured in the AGN frame, unless specified otherwise.

\subsection{Analytical method}
In order to explore the necessity of external photon
fields, the following analytical calculation of searching the parameter space will be carried out under the one-zone SSC model. In the one-zone model, the physical properties of the emitting region are characterized by three parameters, i.e., $R$, $B$, and $\delta$, which can be constrained analytically based on the peak frequencies and peak luminosities of two SED peaks \citep{2017ApJ...842..129C, 2018ApJS..235...39C}. Here we adopt $10^{15}$--$10^{17}$ cm, 0.1--10 G \citep[e.g.,][]{2009MNRAS.400...26O, 2012A&A...545A.113P, 2016A&A...590A..48K, 2017A&A...597A..80H, 2021A&A...651A..74K, 2022MNRAS.510..815K},
and 1--30 \citep{2009A&A...494..527H} as the reference ranges for $R$, $B$, and $\delta$ suggested by observations (hereafter referred to as the observational constraints). If no parameter space could be found within them, introducing the external photon field to explain the high-energy peak would be inevitable.
    
From peak frequencies of the two peaks, the relation between $B$ and $\delta$ can be obtained. In the framework of one-zone SSC model, the emission at the peak frequency is dominantly produced by relativistic electrons with $\gamma_{\rm b}$, where $\gamma_{\rm b}$ is the break Lorentz factor of the electron energy distribution (EED). Using the monochromatic approximation \citep{1979rpa..book.....R}, the peak frequency of low-energy peak $\nu \mathrm{_{s} ^{obs} }$ can be expressed as 
\begin{equation}\label{eq1}
\nu \mathrm{_{s} ^{obs} } =3.7\times 10^{6} \gamma _{\rm b}^{2} B\frac{\delta }{1+z},
\end{equation}
where $z$ is the redshift. In the Thomson (TMS) regime, the peak frequency of the high-energy peak $\nu \mathrm{_{c} ^{obs} }$ is expressed as $\nu \mathrm{_{c} ^{obs} }  =(4/3)\gamma _{\rm b} ^{2} \nu \mathrm{_{s} ^{obs}}$. Combining them together, the relation between $B$ and $\delta$ can be obtained
\begin{equation}\label{eq2}
B=\left ( 1+z \right ) \frac{(\nu \mathrm{_{s} ^{obs} })^{2} }{2.8\times 10^{6}\nu \mathrm{_{c} ^{obs} } } \delta^{-1}.
\end{equation}
It can be seen that when $\nu \mathrm{_s^{obs}}$
and $\nu \mathrm{_c^{obs} }$ are derived from observation, $B$ and $\delta$ are inversely proportional. Since the frequency obtained by the empirical function has some uncertainty, we consider an uncertainty of a factor of 3 in $\nu\mathrm{_{c} ^{obs} }$ in the following.

The relation between $R$, $B$, $\delta$ in the TMS regime can also be obtained from the ratio of the total luminosities of the two peaks,
\begin{equation}\label{eq3}
\frac{L\mathrm{_c^{obs} }}{L\mathrm{_s^{obs} }} =\frac{U_{\rm syn}}{U_{\rm B}},
\end{equation}
where $L\mathrm{_c^{obs} }$ and $L\mathrm{_s^{obs} }$
represent the total luminosities of the IC peak
and the synchrotron peak, respectively;
$U\mathrm{_{syn} } =L\mathrm{_{s} ^{obs} }/(4\pi R^{2} c \delta ^{4})$ and
$U_{\rm B}=B^{2}/(8\pi)$ are the energy densities of synchrotron photons
and magnetic field in the comoving frame, respectively, where $c$ is the speed of light.
If assuming that the shape of two peaks can be represented
by a broken power law spectrum, we have
\begin{equation}\label{eq4}
L\mathrm{^{obs} }  =f\left ( \alpha _{1} ,\alpha _{2}  \right ) L\mathrm{_{p} ^{obs} },
\end{equation}
where $L\mathrm{_{p}^{obs}}$ is the peak luminosity, and $f\left ( \alpha _{1} ,\alpha _{2}  \right )=\frac{1}{\alpha_{2} }-\frac{1}{\alpha_{1}}$ is a correction term, where $\alpha _{1}$ and $\alpha _{2}$ are the slopes below and above the peaks, respectively, in the $\rm log\nu - \rm log\nu\it F_{\nu}$ diagram.
Substituting Eq.~(\ref{eq4}) into Eq.~(\ref{eq3}), we have
\begin{equation}\label{eq5}
B=\left [ \frac{2(L\mathrm{_{s,p}^{obs} )^{2}  } f\left ( \alpha _{1} ,\alpha _{2}  \right )}{L\mathrm{_{c,p}^{obs}  }cR^{2} }  \right ]^{1/2} \delta ^{-2},
\end{equation}
where $L\mathrm{_{s,p}^{obs}}$ and $L\mathrm{_{c,p}^{obs}}$
are the peak luminosities of the low-energy peak and
the high-energy peak, respectively. If considering $R$ satisfies the observational constraint, the correlation between $B$ and $\delta$ can be obtained.
Combining Eq.~(\ref{eq2}) and Eq.~(\ref{eq5}), it is possible to find
that if a reasonable parameter space can be found in one-zone SSC model for LBLs
under the TMS regime.

On the other hand, since the high-energy peaks of LBLs usually extend to $\sim 1~\rm GeV$ band, it is necessary to check if the Klein–Nishina (KN) effect becomes severe and softens the spectrum. When $\gamma _{\rm b}$ and $\nu_{\rm s}$ satisfy
\begin{equation}\label{eq6}
\gamma _{\rm b}\nu_{\rm s}\ge \frac{3}{4}\frac{m_{\mathrm{e} } c ^{2}}{h}, 
\end{equation}
where $h$ is the Planck constant, and
$m_{\mathrm{e}}$ is the electron rest mass, the KN effect will be severe, lowering the peak frequency and
peak luminosity of the IC peak.  Following \citet{1998ApJ...509..608T}, $\nu \mathrm{_{c} ^{obs} }$ in the KN regime can be obtained by
\begin{equation}\label{eq7}
\nu \mathrm{_{c} ^{obs} } =\nu_{\rm c}\frac{\delta }{1+z} \simeq\frac{m_{\mathrm{e}}c^{2}}{h }\gamma _{\rm b}g\left ( \alpha _{1} ,\alpha _{2}  \right )\frac{\delta }{1+z},
\end{equation}
where $g\left(\alpha_{1},\alpha_{2}\right )=\exp\left[\frac{1}{\alpha_{1}}+\frac{1}{2\left(\alpha_{2}-\alpha_{1} \right)}\right]\lesssim 1$. Combining Eq.~(\ref{eq1}) and Eq.~(\ref{eq7}), we can get the relation between $B$ and $\delta$
\begin{equation}\label{eq8}
B=\frac{\nu\mathrm{_s^{obs} } }{(\nu\mathrm{_c^{obs} })^{2}}  \left ( \frac{m_{\mathrm{e} } c^{2} }{h}  \right )^{2} \frac{g\left ( \alpha _{1} ,\alpha _{2}  \right )^{2}}{3.7\times 10^{6} }\frac{1}{1+z}\delta.   
\end{equation}
Interestingly, in the KN regime, the relation between $B$ and $\delta$ turns to be positive,
which indicates that the parameter spaces constrained in the TMS regime
and the KN regime will be quite different.

Due to the severe KN effect, the IC cooling efficiency is greatly reduced,
so the ratio of the total luminosity needs to be corrected as
\begin{equation}\label{eq9}
\frac{L\mathrm{_{c} ^{obs} }  }{L\mathrm{_{s}^{obs}  } }=\frac{U_{\rm syn,avail}}{U_{\rm B}},
\end{equation}
where $U_{\rm syn,avail}$ is the available energy density of synchrotron
photons, which can be obtained by integrating the energy density
$\epsilon  \mathrm{_{syn} }\left ( \nu_{0}  \right )$ of photons
with frequency $\nu_{0}\le 3m_{\mathrm{e} } c^{2} /4h\gamma _{\mathrm{b}}$,
i.e.,
\begin{equation}\label{eq10}
U_{\rm syn,avail}=\int_{0}^{3m_{\mathrm{e} } c^{2} /4h\gamma _{\mathrm{b} } } \epsilon \mathrm{_{syn} }\left ( \nu_{0}  \right )d\nu_{0}=U_{\rm syn}\left ( \frac{3m_{\mathrm{e} } c^{2} \delta }{4h\gamma _{\mathrm{b} } \nu \mathrm{_{s} ^{obs} }  }  \right ) ^{-\alpha_{1}}.
\end{equation}
Substituting Eq.~(\ref{eq4}), Eq.~(\ref{eq7}), and Eq.~(\ref{eq10})
into Eq.~(\ref{eq9}),
we can obtain the relation between $R$, $B$, and $\delta$ in the KN regime as
\begin{equation}\label{eq11}
B=\left \{ \frac{2(L \mathrm{_{s,p}^{obs}  } )^{2}f\left ( \alpha _{1} ,\alpha _{2}  \right ) }{R^{2} c\delta ^{4+2\alpha _{1} } L \mathrm{_{c,p}^{obs}  }\left [ \frac{3}{4}\left ( \frac{m_{\mathrm{e} } c^{2} }{h} \right )^{2} \frac{g\left ( \alpha _{1} ,\alpha _{2}  \right )}{\nu \mathrm{_{s} ^{obs} }  \nu \mathrm{_{c} ^{obs} }\left ( 1+z \right )  } \right ]^{\alpha_{1}}   }  \right \} ^{1/2}.  
\end{equation}

\begin{figure}
\centering
\subfloat{\includegraphics[width=1\columnwidth]{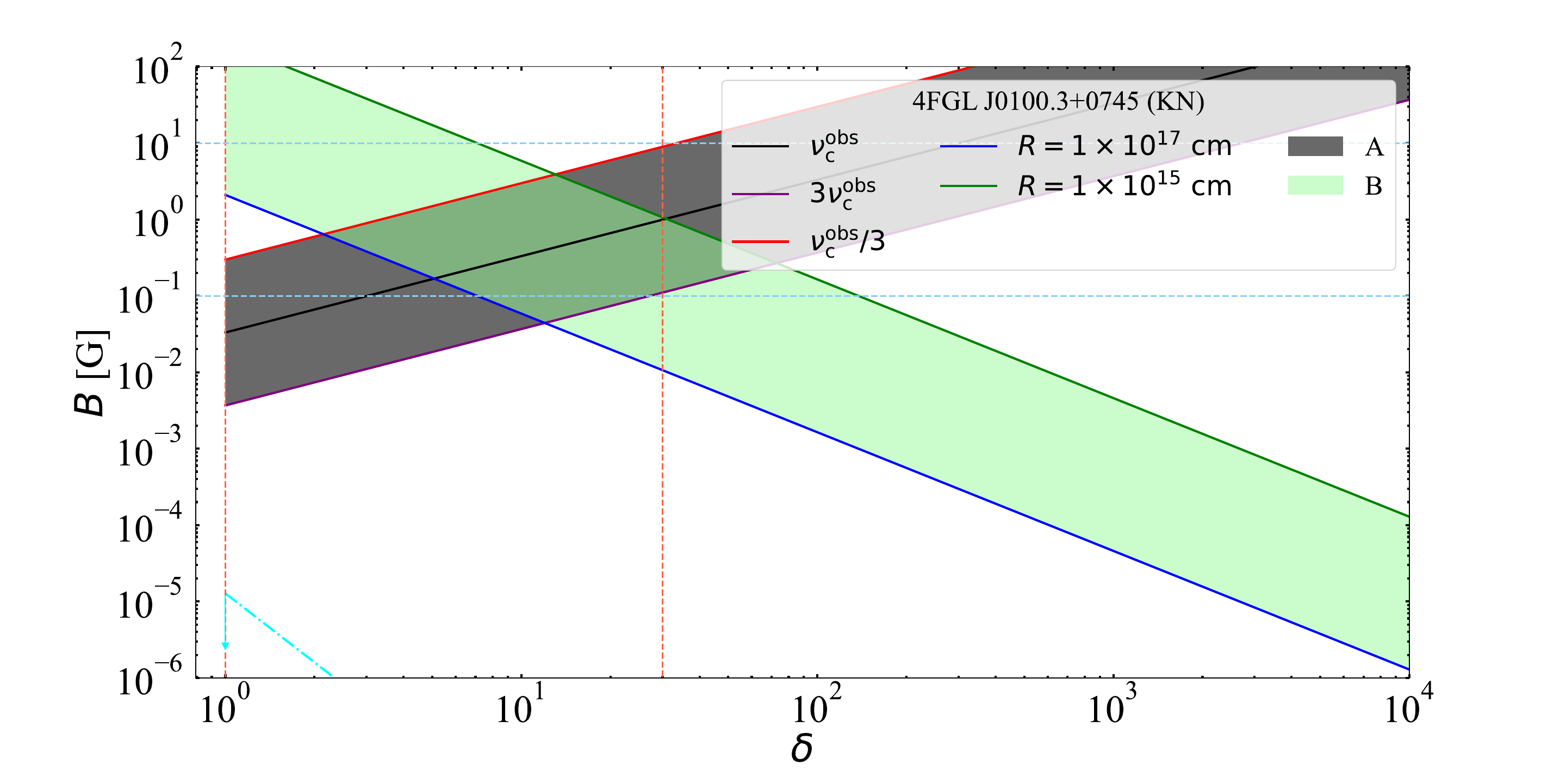}}
\hspace{-5mm}
\subfloat{\includegraphics[width=1\columnwidth]{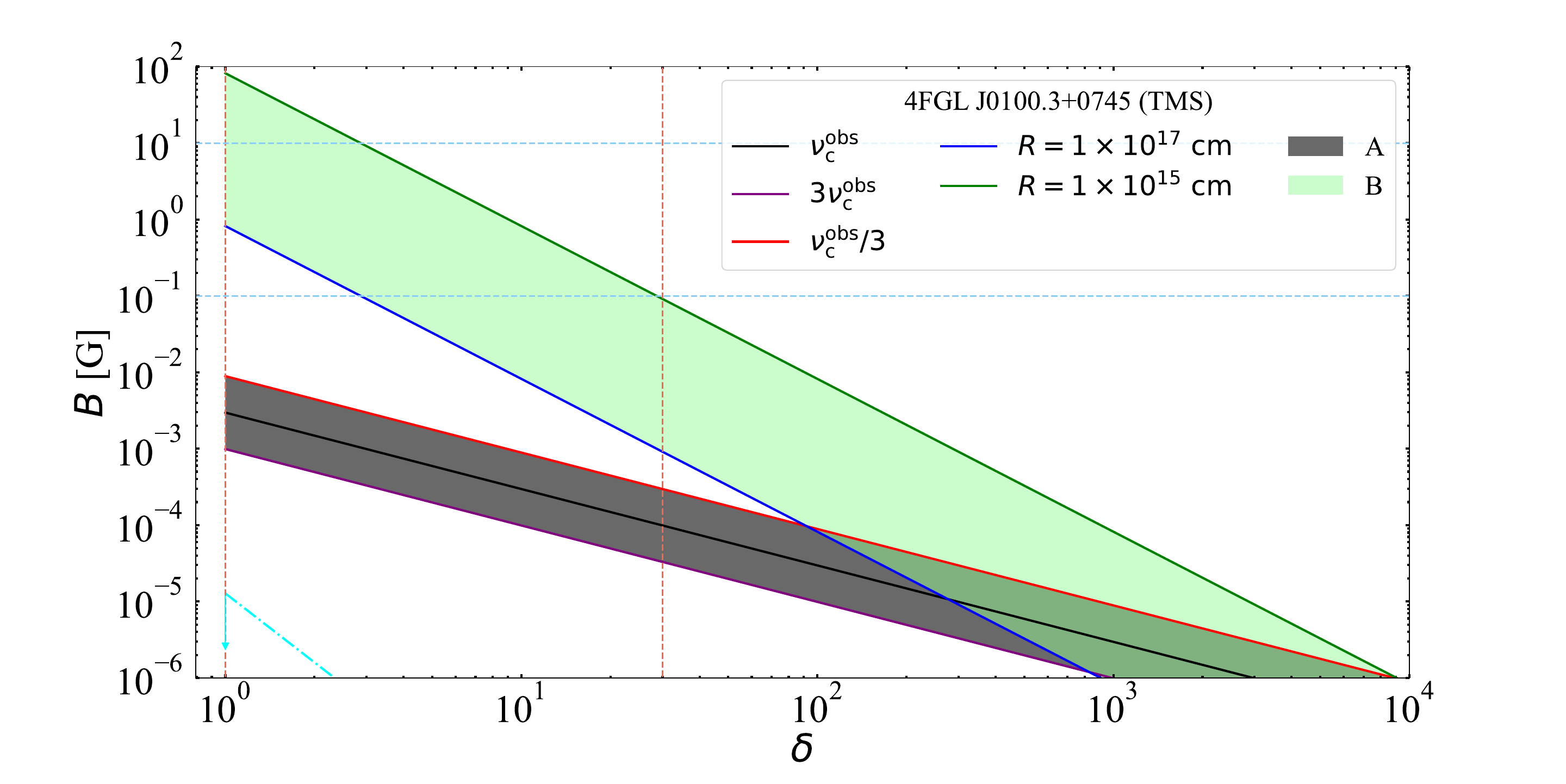}}
\caption{The parameter spaces of 4FGL J0100.3+0745, with the upper panel for
the KN regime and the lower panel for the TMS regime.
The gray region A represents the peak frequency constraint, corresponding to
Eq.~(\ref{eq8}) and Eq.~(\ref{eq2}), respectively.
The palegreen region B represents the peak luminosity constraint,
corresponding to Eq.~(\ref{eq11}) and Eq.~(\ref{eq5}), respectively.
The red vertical and blue horizontal dashed lines represent
the reference range of $\delta$ and $B$ suggested by observations, respectively.
The cyan dash-dotted line with arrows represents the parameter space that
the severe KN effect is trigged, corresponding to Eq.~(\ref{eq12}).
The meaning of other line styles are given in the legends.}
\label{fig1}
\end{figure}

Finally, we can analyse the applicability of the one-zone SSC model to LBLs in
different scattering regimes.
Here, we take the first LBL of our sample, i.e., 4FGL J0100.3+0745, as an example to show its parameter space derived with above methods. The resulting parameter spaces are shown in
Fig.~\ref{fig1}, where the upper panel is for the KN regime,
and the lower panel is for the TMS regime.
Combining Eq.~(\ref{eq1}) and Eq.~(\ref{eq6}),
we can find that only when
\begin{equation}\label{eq12}
B\le\frac{16\left ( 1+z \right )^{3} h^{2} (\nu \mathrm{_{s} ^{obs} } )^{3}  }{9\times 3.7\times 10^{6} m_{\mathrm{e} } ^{2} c^{4} } \delta ^{-3},   
\end{equation}
the severe KN effect will be triggered.
Therefore, the space above the cyan dash-dotted line
in the upper panel of Fig.~\ref{fig1} should be discarded.
Since no parameter space can be found under the observational constraints
(corresponding to the rectangular region), we suggest that
its SED cannot be fitted with the one-zone SSC model in the KN regime.
To explore the possibility of reproducing SEDs in the KN regime for general cases,
we give a test in the critical condition ($\delta=1$, $B=0.1~\mathrm{G}$).
By setting $z=0.27$
\citep[an average value suggested by][]{2011ApJ...743..171A},
we find that only when $\nu \mathrm{_{s} ^{obs} } \ge 1.16\times 10^{15}$ Hz,
there is a intersect area between the cyan dash-dotted line
and the rectangular region.
Therefore, for LBLs, we only need to consider whether there is a parameter space
satisfying observational constraints in the TMS regime.
However, it can be seen from the lower panel of Fig.~\ref{fig1}
that there is still no parameter space.
Therefore, the external photon field is necessary to be introduced.
The above analytical methods would be applied in this work
and the corresponding results would be given in Section~\ref{result}.

\subsection{Model description}
The external photon fields in the mostly applied one-zone EC model are BLR and DT, which are closely related to the thermal radiation
of the accretion disk. In our modeling, we assume that the accretion disk is geometrically thin and
optically thick \citep{1973A&A....24..337S}.
Its emission profile has the following multi-temperature
blackbody form \citep{2002apa..book.....F}:
\begin{equation}\label{eq13}
F\mathrm{^{obs}_{disk}}(\nu\mathrm{^{obs}})=(\nu\mathrm{^{obs} })^{3}\frac{4\pi h\cos\theta\mathrm{^{obs} }}{c^{2}(D\mathrm{_{L}^{obs})}^{2}}\int_{R\mathrm{_{disk,in}}}^{R\mathrm{_{disk,out}}}\frac{R\mathrm{_{disk}}dR\mathrm{_{disk}}}{e^{h\nu\mathrm{^{obs}}/kT(R\mathrm{_{disk}})}-1},
\end{equation}
where $D\mathrm{_{L}^{obs}}$ is the luminosity distance, $k$ is the
Boltzmann constant, $R\mathrm{_{disk}}$ is the radial distance between
the disk and the central supermassive black hole (SMBH),
$R\mathrm{_{disk,in}}$ and $R\mathrm{_{disk,out}}$ are the inner and outer
radii of the disk, assumed to be $3R\mathrm{_{Sch}}$ and
$500R\mathrm{_{Sch}}$, respectively \citep[e.g.,][]{2002apa..book.....F},
where $R\mathrm{_{Sch}}=2GM\mathrm{_{BH}}/c^{2}$ is the Schwarzschild radius,
$G$ is the universal gravitational constant, $M\mathrm{_{BH}}$ is the mass
of the SMBH.
Note that, $M\mathrm{_{BH}}$ only affects the peak frequency of
the thermal radiation of the accretion disk. For simplicity, $M\mathrm{_{BH}}$ is assumed to be an average value of $10^{9}$ solar mass
\citep{2017ApJ...851...33P, 2022ApJ...925...40X}.
The radial dependence of the temperature is as follows:
\begin{equation}\label{eq14}
T(R\mathrm{_{disk}})=\left\{\frac{3R\mathrm{_{Sch}}L\mathrm{_{disk }^{AGN}}}{16\pi\eta\mathrm{_{acc}}\sigma\mathrm{_{SB}}(R\mathrm{_{disk}})^{3}}\left [ 1-\left(\frac{3R\mathrm{_{Sch} }}{R\mathrm{_{disk}}}\right)^{1/2}\right ]\right\}^{1/4},
\end{equation}
where $L\mathrm{_{disk }^{AGN}}$ is the luminosity of the accretion disk,
$\sigma\mathrm{_{SB}}$ is the Stefan-Boltzmann constant,
$\eta\mathrm{_{acc}}=10\%$ is the accretion efficiency \citep[e.g.,][]{2017ApJ...851...33P}.
Since the emitting region is moving away from the accretion disk,
the energy density of rear-end photons from the disk is usually low
in the comoving frame. Therefore, in this work, we mainly consider
the BLR and DT as the external photon fields
\citep[e.g.,][]{2016MNRAS.463.4469D, 2018A&A...616A..63A},
which reprocess $\eta\mathrm{_{BLR}}=\eta\mathrm{_{DT}}=0.1$ of the disk
luminosity into the BLR and DT radiation, respectively
\citep[e.g.,][]{2012ApJ...754..114H}.
The energy densities of the BLR ($U\mathrm{_{BLR}}$) and the DT ($U\mathrm{_{DT}}$)
in the comoving frame can be estimated as \citep{2012ApJ...754..114H}
\begin{equation}\label{eq15}
U\mathrm{_{BLR}} = \frac{\eta \mathrm { _ { BLR } } \Gamma ^ { 2 } L \mathrm { _ { disk } ^ { AGN } } } { 4 \pi ( r \mathrm { _ { BLR } ^ { AGN } } ) ^ { 2 } c \left[ 1 + \left(r\mathrm{^{ AGN }}/r\mathrm {_{ BLR}^{AGN}}\right)^{3}\right]}   
\end{equation}
and
\begin{equation}\label{eq16}
U\mathrm{_{DT}}=\frac{\eta\mathrm{_{ DT }}\Gamma^{ 2 }L\mathrm{_{ disk}^{ AGN}}}{ 4 \pi ( r \mathrm { _ { DT } ^ { AGN } } ) ^ { 2 } c \left[ 1 + \left( r \mathrm { ^ { AGN } } / r \mathrm { _ { DT } ^ { AGN } } \right) ^ { 4 } \right] },   
\end{equation}
where $r\mathrm{^{AGN}}$ is the distance between the emitting region and the SMBH,
$r\mathrm{^{AGN}_{BLR}}=0.1(L\mathrm{_{disk}^{AGN}}/10^{46}\mathrm{~erg~s^{-1}})^{1/2}$
pc and
$r\mathrm{^{AGN}_{DT}}=2.5(L\mathrm{_{disk}^{AGN}}/10^{46}\mathrm{~erg~s^{-1}})^{1/2}$
pc are the characteristic distances of the BLR and DT, respectively 
\citep[e.g.,][]{2008MNRAS.387.1669G}.
The radiation of the BLR and DT are assumed to be isotropic graybody radiation,
whose peak frequencies in the comoving frame are $2\times10^{15}\Gamma$ Hz
\citep{2008MNRAS.386..945T}
and $ 3\times10^{13}\Gamma$ Hz 
\citep{2007ApJ...660..117C}, respectively.

In the leptonic model, the jet's non-thermal emission is from the relativistic elections, whose distribution can be obtained
by solving the continuity equation that includes injection,
radiative cooling and escape
\citep[e.g.,][]{1999MNRAS.306..551C, 2023ApJ...948...82H}.
In this work, relativistic electrons are assumed to be injected into
the emitting region with a broken power-law distribution at a constant rate
\citep{2010MNRAS.402..497G}, i.e.,
\begin{equation}\label{eq17}
Q(\gamma)=Q_{0}\gamma^{-p_{1}} \left[1+\left (\frac{\gamma }{\gamma\mathrm{_{b}}} \right)^{p_{2}-p_{1}}\right]^{-1},~\gamma\mathrm{_{min}}<\gamma\mathrm{_{b}}<\gamma\mathrm{_{max}},
\end{equation}
where $\gamma$ is the electron Lorentz factor, $\gamma\mathrm{_{min}}$ and
$\gamma\mathrm{_{max}}$ are the minimum and maximum electron Lorentz factors,
respectively, $p_{1}$ and $p_{2}$ are the spectral indices below and above
$\gamma\mathrm{_{b}}$, respectively, $Q_{0}$ is a normalization constant
in units of $\mathrm{s}^{-1}~\mathrm{cm}^{-3}$,
which can be obtained from $\int Q(\gamma)\gamma m\mathrm{_{e}}c^{2}d\gamma=\frac{3L\mathrm{_{e,inj}}}{4\pi R^{3}}$,
where $L\mathrm{_{e,inj}}$ is the electron injection luminosity. 
When the injection is balanced by the cooling and escape \citep{2021MNRAS.506.5764D, 1999MNRAS.306..551C, 2000ApJ...536..729L, 2002ApJ...581..127B, 2011ApJ...739...66T, 2012MNRAS.424.2173Y},
the steady-state EED $N(\gamma)$ is obtained as $N(\gamma)\approx Q(\gamma)t\mathrm{_{e}},$ where
$t\mathrm{_{e}}=\mathrm{min}\left\{t\mathrm{_{dyn}},~t\mathrm{_{cool}}\right\}$. More specifically, $t\mathrm{_{dyn}}=R/c$ is the dynamical timescale, and
$t\mathrm{_{cool}}=3m\mathrm{_{e}}c/\left [4(U\mathrm{_{B}}+\kappa \mathrm{_{KN}}U\mathrm{_{ph}})\sigma \mathrm{_{T}}\gamma\right]$
is the radiative cooling timescale,
where $\sigma\mathrm{_{T}}$ is the Thomson scattering cross-section,
\begin{equation}\label{eq18}
\begin{aligned}
\kappa \mathrm{_{KN}}=&\frac{9}{U\mathrm{_{ph}}}\int_{0}^{\infty}dEEn\mathrm{_{ph} }(E)\\
&\int_{0}^{1}dq\frac{2q^{2}\ln{q}+q(1+2q)(1-q)+\frac{q(\omega q)^{2}(1-q)}{2(1+\omega q)}}{(1+\omega q)^{3}}
\end{aligned}
\end{equation}
is a numerical factor accounting for the KN effect
\citep[e.g.,][]{2010NJPh...12c3044S, 2022PhRvD.106j3021X},
where $E$ is the energy of soft photons,
$\omega=4E\gamma/(m\mathrm{_{e}}c^{2})$,
$n\mathrm{_{ph}}(E)$ is
the number density distribution of soft photons, 
$U\mathrm{_{ph}}=U\mathrm{_{syn}} +U\mathrm{_{ext}}$ is the energy density of
soft photons. For SSC model, we set $U\mathrm{_{ext}}=0$, while $U\mathrm{_{ext}}=U\mathrm{_{BLR}} +U\mathrm{_{DT}}$ for EC model.

After obtaining the steady-state EED $N(\gamma)$, we can calculate
the non-thermal radiation of the jet, including synchrotron, SSC, and EC emissions. For the synchrotron emission, its emission coefficient can be obtained by
\begin{equation}\label{eq19}
j\mathrm{_{syn} } (\nu)=\frac{1}{4\pi}\int N(\gamma)P(\nu,\gamma)d\gamma,
\end{equation}
where $P(\nu,\gamma)$ is the mean emission coefficient for a single
electron integrated over the isotropic distribution of pitch angles
\citep[e.g.,][]{1986A&A...164L..16C, 1988ApJ...334L...5G, 2001A&A...367..809K}.
And the synchrotron absorption coefficient
\citep[e.g.,][]{1979AstQ....3..199R} can be calculated with
\begin{equation}\label{eqdd20}
k\mathrm{_{syn}}(\nu)=-\frac{1}{8\pi m\mathrm{_{e}}\nu^{2}}\int\gamma^{2}\frac{\partial}{\partial\gamma}\left[\frac{N(\gamma)}{\gamma^{2}}\right]P(\nu,\gamma)d\gamma.
\end{equation}
Then, we can calculate the synchrotron intensity by solving
radiative transfer equation for the spherical geometry
\citep[e.g.,][]{1996ApJ...461..657B, 1999ApJ...514..138K}:
\begin{equation}\label{eqdd21}
I\mathrm{_{syn}}(\nu)=\frac{j\mathrm{_{syn}}(\nu)}{k\mathrm{_{syn}}(\nu)}\left[1-\frac{2}{\tau(\nu)^{2}}(1-\tau(\nu)\mathrm{e}^{-\tau(\nu)}-\mathrm{e}^{-\tau(\nu)})\right], 
\end{equation}
where $\tau(\nu)=2Rk\mathrm{_{syn}}(\nu)$ is the optical depth.

The SSC and EC emission coefficients are given as
\begin{equation}\label{eq20}
j\mathrm{_{IC}}(\nu)=\frac{h\epsilon }{4\pi}\int d\epsilon_{0} n(\epsilon_{0})\int d\gamma N(\gamma)C(\epsilon,\gamma,\epsilon_{0}),
\end{equation}
where $\epsilon_{0}$ and $\epsilon$ are the soft photon energy and the
scattered photon energy in units of $m\mathrm{_{e}}c^{2}$, respectively,
$n(\epsilon_{0})$ is the number density of soft photons per energy interval,
$C(\epsilon,\gamma,\epsilon_{0})$ is the Compton kernel given by
\citet{1968PhRv..167.1159J}.
Since the emitting region is transparent for IC radiation, we can obtain the IC intensity as
$I\mathrm{_{IC}}(\nu)=j\mathrm{_{IC}}(\nu)R.$ Finally, the total observed
flux density of the jet can be calculated by
\begin{equation}\label{eq21}
F\mathrm{^{obs}_{jet}}(\nu\mathrm{^{obs}})=\frac{\pi R^{2}\delta^{3}(1+z)}{(D\mathrm{_{L}^{obs}})^{2}}\left (I\mathrm{_{syn}}(\nu)+I\mathrm{_{IC}}(\nu)\right).
\end{equation}
Due to the possibility of $\gamma\gamma$ annihilation between
high-energy photons and low-energy photons, we calculate
the internal $\gamma\gamma$ absorption \citep[e.g.,][]{2009herb.book.....D} and correct the GeV–TeV spectrum
by using the extragalactic background light model presented by
\citet{2021MNRAS.507.5144S}.

There are 11 free parameters in our model: $\delta$, $B$, $R$, $L\mathrm{_{e,inj}}$,
$\gamma\mathrm{_{b}}$, $p_{1}$, $p_{2}$, $\gamma\mathrm{_{min}}$,
$\gamma\mathrm{_{max}}$, $L\mathrm{_{disk}^{AGN}}$, and $r\mathrm{^{AGN}}$ only for
the EC model. Since they are coupled to each other, reproducing the best-fit SEDs
will take a long time if we allow all parameters to be free.
In this work, we estimate $p_{1}$ and $p_{2}$ from the spectral indices
$\alpha_{1}$ and $\alpha_{2}$ of the SEDs, respectively.
We set $\gamma\mathrm{_{min}}=1$, because it has little effect
on the fitting results. We adopt $\gamma\mathrm{_{max}}=2\times10^{6}$
\citep{2005A&A...432..401G}
as a  default value, unless it is constrained by the quasi-simultaneous data
of the low-energy peak \citep[e.g.,][]{2012ApJ...752..157Z}.
Finally, we determine $L\mathrm{_{disk}^{AGN}}$ by fitting the optical-UV data if there is a blue bump structure, otherwise we assume that $L\mathrm{_{disk}^{AGN}}$
can be any value that does not contaminate the synchrotron emission.

\begin{table*}
\scriptsize 
\caption{The parameters used to fit the SEDs. Columns from left to right: (1) source
name; (2) Doppler factor; (3) magnetic field in units of Gauss; (4) radius of the
emitting region in units of cm; (5) electron injection luminosity in units of
$\mathrm{erg}~\mathrm{s}^{-1}$; (6) break electron Lorentz factor; (7) and (8) are
the spectral indices below and above $\gamma\mathrm{_{b}}$, respectively;
(9) maximum electron Lorentz factor; (10) accretion disk luminosity in units of
$\mathrm{erg}~\mathrm{s}^{-1}$; (11) distance between the emitting region and the SMBH
in units of pc; (12) chi-square value, $\chi^{2}=\frac{1}{m-\mathrm{dof}}{\textstyle\sum_{i=1}^{m}}\left(\frac{\hat{y}_{i}-y_{i}}{\sigma_{i}}\right)^{2}$,
where $m$ is the number of quasi-simultaneous observational data points, dof are the
degrees of freedom, $\hat{y}_{i}$ are the expected values from the model,
$y_{i}$ are the observed data, and $\sigma_{i}$ is the
standard deviation for each data point.
In our sample, the errors of the data points from infrared to X-ray bands are
collected from the SSDC website, while the errors of the $\gamma$-ray data points are
obtained by the $Fermi$-LAT data analysis
(details can be found in Section~\ref{fermi}).
}\label{table:parameters}
\centering
\begin{threeparttable}
\begin{tabular}{cccccccccccc}
\hline\hline
Source name & $\delta$ & $B$ & $\log{R}$ & $\log{L\mathrm{_{e,inj}}}$ & $\log{\gamma\mathrm{_{b}}}$ & $p_{1}$ & $p_{2}$ & $\log{\gamma\mathrm{_{max}}}$ & $\log{L\mathrm{_{disk}^{AGN}}}$ & $\log{r\mathrm{^{AGN}}}$ & $\chi^{2}$\\
& & (G) & (cm) & ($\mathrm{erg~s^{-1}}$) & & & & & ($\mathrm{erg~s^{-1}}$) & (pc) &\\
~(1) & (2) & (3) & (4) & (5) & (6) & (7) & (8) & (9) & (10) & (11) & (12)\\
\hline
4FGL J0100.3+0745 & 29.0 & 1.0 & 16.8 & 39.8 & 2.7 & 0.1 & 3.1 & 6.3 & 42.5 & -2.2 & 6.1\\
4FGL J0141.4-0928\tnote{a} & 9.7 & 1.0 & 16.7 & 43.3 & 2.6 & 1.0 & 2.5 & 4.0 & 43.9 & -1.5 & 7.5\\
4FGL J0210.7-5101\tnote{a} & 18.5 & 2.0 & 16.4 & 42.8 & 2.8 & 1.2 & 3.9 & 3.2 & 46.2 & -0.4 & 13.4\\
4FGL J0238.6+1637\tnote{ab} & 15.1 & 0.8 & 16.6 & 43.0 & 2.7 & 0.7 & 3.7 & 3.6 & 43.9 & -1.4 & 141.6\\
4FGL J0334.2-4008\tnote{ab} & 26.0 & 0.5 & 16.9 & 42.8 & 2.8 & 1.5 & 2.9 & 3.8 & 43.9 & -0.8 & 34.3\\
4FGL J0854.8+2006\tnote{ab} & 11.7 & 0.9 & 16.9 & 42.9 & 2.7 & 1.1 & 2.6 & 4.3 & 43.9 & -1.2 & 82.2\\
4FGL J0958.7+6534\tnote{b} & 16.0 & 1.0 & 16.6 & 42.3 & 2.8 & 0.9 & 4.2 & 6.3 & 43.9 & -1.3 & 36.4\\
4FGL J1043.2+2408\tnote{b} & 16.0 & 1.5 & 16.4 & 42.3 & 2.6 & 0.7 & 4.5 & 6.3 & 45.2 & -0.9 & 14.4\\
4FGL J1147.0-3812\tnote{a} & 25.8 & 0.3 & 16.9 & 42.9 & 2.2 & 1.2 & 3.0 & 4.0 & 43.5 & -0.8 & 19.6\\
4FGL J1517.7-2422\tnote{a} & 4.8 & 0.3 & 16.6 & 42.9 & 3.4 & 1.8 & 2.9 & 4.9 & 43.0 & -1.6 & 35.9\\
4FGL J1751.5+0938\tnote{b} & 11.7 & 0.3 & 17.0 & 43.3 & 2.6 & 1.8 & 3.2 & 6.3 & 43.7 & -1.1 & 68.9\\
4FGL J1800.6+7828\tnote{a} & 17.9 & 1.0 & 16.6 & 42.7 & 3.3 & 1.6 & 3.7 & 4.2 & 43.2 & -1.6 & 8.9\\
4FGL J2152.5+1737 & 26.3 & 0.6 & 16.7 & 42.0 & 3.3 & 1.5 & 4.7 & 6.3 & 43.9 & -1.1 & 14.5\\
4FGL J2247.4-0001 & 16.5 & 0.5 & 16.7 & 42.7 & 3.4 & 1.7 & 3.6 & 6.3 & 43.9 & -1.3 & 5.3\\
\hline
4FGL J0522.9-3628\tnote{b} (EC) & 5.0 & 0.3 & 16.8 & 43.1 & 3.1 & 1.5 & 3.1 & 6.3 & 43.0 & -1.5 & 68.6\\
4FGL J0522.9-3628\tnote{b} (SSC) & 6.2 & 0.1 & 17.0 & 43.3 & 3.4 & 1.6 & 3.8 & 6.3 & 43.0 & & 82.8\\

\hline
\end{tabular}
\begin{tablenotes}
\footnotesize
\item[\textbf{Notes.}]
\item[a] The source with limited $\gamma\mathrm{_{max}}$.
\item[b] The source with potentially more than one state of
quasi-simultaneous data points in the optical and/or X-ray bands.
\end{tablenotes}
\end{threeparttable}
\end{table*}

\section{Results and discussion}\label{result}
\subsection{The physical properties of LBLs}\label{result4.1}
By applying the analytical method to all 15 LBLs in our sample,
we find that except for 4FGL J0522.9-3628, the quasi-simultaneous multi-wavelength
SEDs of the rest sources cannot be fitted with the one-zone SSC model.
This indicates the importance of external photon fields for the high-energy emission of LBLs.
Combining the observational feature that BL Lacs lack strong emission lines,
our results suggest that LBLs may be the masquerading BL Lacs,
whose broad emission lines are outshone by the non-thermal radiation of the jet 
\citep{2013MNRAS.431.1914G, 2019MNRAS.484L.104P, 2022ApJ...936..146X}.
The fitting results with the one-zone EC model are shown
in Fig.~\ref{fig2}, and the fitting parameters are presented
in the upper part of Table~\ref{table:parameters}.
For the modeling of 4FGL J0522.9-3628, the upper panel of Fig.~\ref{fig3}
shows its fitting result with the conventional one-zone EC model,
while the middle panel shows the parameter space of the one-zone SSC model in the TMS regime.
As indicated by the derived parameter space, we fit the SED with the one-zone SSC model. The fitting result is shown
in the lower panel of Fig.~\ref{fig3}, and the parameters used for fitting
are presented in the lower part of Table~\ref{table:parameters},
where $\delta$ and $B$ are selected from the middle panel (corresponding to the red cross).
It can be seen that the one-zone SSC model can fit the SED well
even without external photon fields.
However, this is a rare occurrence (only 1 out of 15 sources in our sample).

Due to the constraint of quasi-simultaneous observational data
(see Fig.~\ref{fig2}), as shown in Table~\ref{table:parameters},
the sources with superscript "a"
have $\log{\gamma\mathrm{_{max}}}$ less than 6.3
(corresponding to $\gamma\mathrm{_{max}} =2\times10^{6}$),
which implies relatively slow shock speeds.
In the framework of diffusive shock acceleration mechanism, 
by equating $t\mathrm{_{e}}$ (corresponding to $\gamma\mathrm{_{max}}$)
with the acceleration timescale
\citep[e.g.,][]{2007Ap&SS.309..119R, 2019ApJ...886...23X},
we can estimate the shock speed measured in the upstream frame as
$u\mathrm{_{s}}=\left[6\gamma\mathrm{_{max}}m\mathrm{_{e}}c^{3}/(eBt\mathrm{_{e}})\right]^{1/2}$.
Then, we find that $\log{u\mathrm{_{s}}}$ in units of $c$ for the sources with
superscript "a" in Table~\ref{table:parameters} are -3.5, -4.2, -4.0, -3.8, -3.4, -3.7,
-2.9, and -3.4, from top to bottom, respectively.
On the other hand, as can be seen from Table~\ref{table:parameters}, sources marked with a superscript "b" have relative large $\chi^2$ values, which are due to the fact that their collected optical and/or X-ray bands quasi-simultaneous data points are rather scattered. It indicates that the quasi-simultaneous data collected from these sources within a week exhibits multiple states, including at least one flare. In the modeling, we selected the ones with shorter error bars for fitting throughout this work, which are more creditable.

The location of the emitting region is one of the essential properties in blazars,
which can be explored by investigating the dominant ambient photon fields
for the EC process
\citep[e.g.,][]{2011ApJ...726L..13A, 2012ApJ...758L..15D, 2014ApJ...789..161N, 2016ApJ...821..102B}.
However, the location of the blazar $\gamma$-ray emitting region is still controversial \citep{2016ARA&A..54..725M}.
Following previous studies
\citep[e.g.,][]{2009ApJ...692...32D, 2009MNRAS.397..985G, 2012arXiv1202.6193G, 2012MNRAS.421.2956Z, 2018ApJ...859..168Y},
we constrain the location of the emitting region by fitting
the quasi-simultaneous SEDs of 15 LBLs in $Fermi$-4LAC with the one-zone leptonic model.
The distance ($r\mathrm{^{AGN}}$) between the emitting region and the SMBH is presented
in Table~\ref{table:parameters}, which is determined by combining Eq.~(\ref{eq15})
and Eq.~(\ref{eq16}) to reproduce the $\gamma$-ray spectrum based on the assumption that
the energy density of the external photon fields are functions of $r\mathrm{^{AGN}}$.
In the upper panel of Fig.~\ref{fig4}, we plot $r\mathrm{^{AGN}}$ as a function of
$L\mathrm{_{disk}^{AGN}}$ for all 15 sources in our sample. It can be seen that most
LBLs have their $\gamma$-ray emitting regions located outside the BLR and within the DT.
On the other hand, it can be found that the soft photons for the EC radiation of 8 LBLs
are dominated by that from the BLR and the soft photons of other 7 LBLs are dominated
by that from DT (see Fig.~\ref{fig2} and Fig.~\ref{fig3}).
Therefore, our modeling results suggest that the $\gamma$-ray emitting regions of LBLs are located inside the DT and close to the characteristic distance of BLR.

To test the above results on the location of the emitting region,
it might be possible to use the variability timescale to diagnose the emitting region as well
\citep[e.g.,][]{2010ApJ...712..957A, 2010ApJ...715..362J, 2010MNRAS.405L..94T, 2011MNRAS.418...90L, 2012JPhCS.355a2032A, 2012ApJ...751L...3G, 2013MNRAS.431..824B, 2015MNRAS.452.1280R}.
Based on the causality relation, we have $R\approx ct\mathrm{_{var}^{obs}}\delta/(1+z)$,
where $t\mathrm{_{var}^{obs}}$ is the variability timescale.
On the other hand, in the framework of the conical jet model,
we have $r\mathrm{^{AGN}}=R/\tan\psi\mathrm{^{AGN}}$,
where $\psi\mathrm{^{AGN}}$ is the semi-aperture angle, assumed to be 0.1 for all sources
\citep[e.g.,][]{2010MNRAS.402..497G}.
By setting $\delta=10$
\citep[a typical value suggested by observations;][]{ 2009A&A...494..527H},
and taking advantage of the variability timescale
(the minimum $\gamma$-ray variability timescales for some individual objects in our sample)
from \citet{2013ApJ...767..103V}, we find that
$\log{r\mathrm{^{AGN}}}$ in units of pc for 4FGL J0210.7-5101, 4FGL J0522.9-3628,
4FGL J1751.5+0938, and 4FGL J1800.6+7828 are -0.7, -1.5, -0.7, and -1.3, respectively.
The distribution of their emitting region locations
is shown in the middle panel of Fig.~\ref{fig4}.
Since the thermal radiation of the accretion disk may be outshone by the luminous
non-thermal radiation from the jet, the adopted value of $L\mathrm{_{disk}^{AGN}}$ can affect the
location of the emitting region.
Here we take 4FGL J1751.5+0938 (this source is located inside the DT
in the upper panel of Fig.~\ref{fig4} and is marked as a green point in the middle
panel of Fig.~\ref{fig4}) as an example to show the relation between
$r\mathrm{^{AGN}}$ and $L\mathrm{^{AGN}_{disk}}$.
The corresponding result is shown in the lower panel of Fig.~\ref{fig4}.
It can be found that, when $U\mathrm{_{ext}}$ remains unchanged,
there is a positive correlation between $r\mathrm{^{AGN}}$ and $L\mathrm{^{AGN}_{disk}}$.
Therefore, if the variability timescale corresponds to the SED we collected,
the difference in the distribution of 4FGL J1751.5+0938 in the upper panel and the middle
panel may indicate an underestimation of $L\mathrm{^{AGN}_{disk}}$.
And we suggest that the $\gamma$-ray emitting regions of LBLs
are still located outside the BLR and within the DT,
i.e., the conclusion should remain unchanged.

The distribution of the emitting region locations may also be roughly explained
from the perspective of the KN effect. More specifically, since the soft photons
from BLR have higher energy, the corresponding EC emission
will be suppressed by the KN effect earlier
\citep[e.g.,][]{2013MNRAS.436.2170C},
leading to a steep spectrum when
$\nu\mathrm{^{obs}}\gtrsim(1/12)(m\mathrm{_{e}}c^{2}/h)\left[m\mathrm{_{e}}c^{2}/(h\nu\mathrm{_{soft}^{AGN}})\right]\approx6\times10^{23}~\mathrm{Hz}$, where $\nu\mathrm{_{soft}^{AGN}}$ is the peak frequency of the external photon field emission. 
However, if the soft photons come from DT, we have
$\nu\mathrm{^{obs}}\gtrsim4\times10^{25}~\mathrm{Hz}$,
which is normally not important for LSPs.
Due to the lack of KN features in the broadband spectra of bright blazars
\citep[e.g.,][]{2009ApJ...704...38S},
the emitting region needs to be outside the BLR.

Our results are consistent with many previous works. For example,
\citet{2020ApJS..248...27T} modelled the quasi-simultaneous SEDs of 60 $Fermi$-4LAC FSRQs
and suggested that most of the $\gamma$-ray emitting regions
are located outside the BLR and within the DT.
\citet{2011ApJ...735..108C} proposed that the IR external photon field may play
an important role by analyzing the ratio of EC to synchrotron luminosity for a sample of
$Fermi$ bright blazars, implying that the emitting regions should be
outside the BLR and within the DT.
Moreover, if the $\gamma$-ray emitting regions are inside the BLR,
a feature of cutoff compatible with the $\gamma\gamma$ interaction with BLR photons
is expected in the high-energy band.
However, only 10\% of the broad-line blazars
show the matching attenuation \citep{2018MNRAS.477.4749C}.
Therefore, the $\gamma$-ray emission should be produced outside the BLR most of the time.

\onecolumn
\begin{figure}
\centering
\includegraphics[width=8cm,height=5cm]{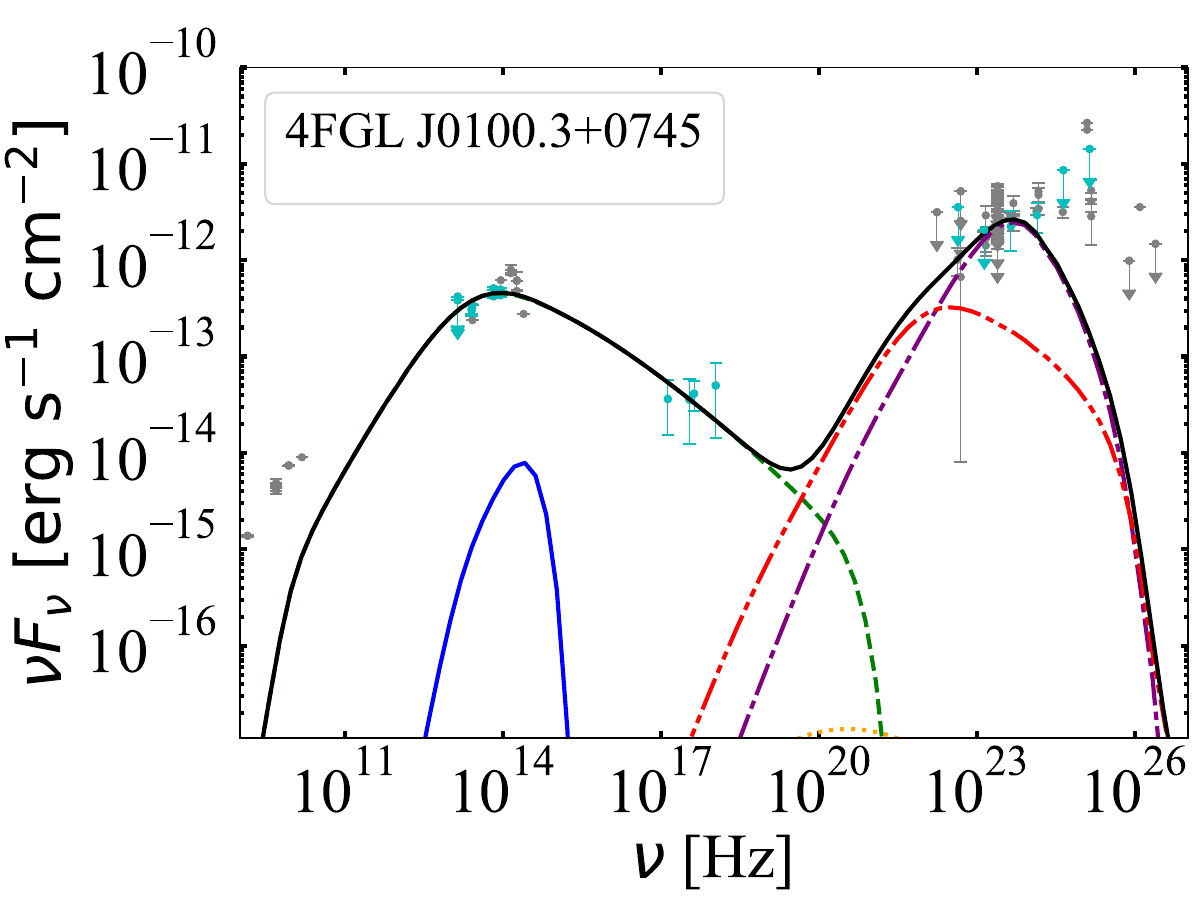}
\hspace{1.2cm}
\includegraphics[width=8cm,height=5cm]{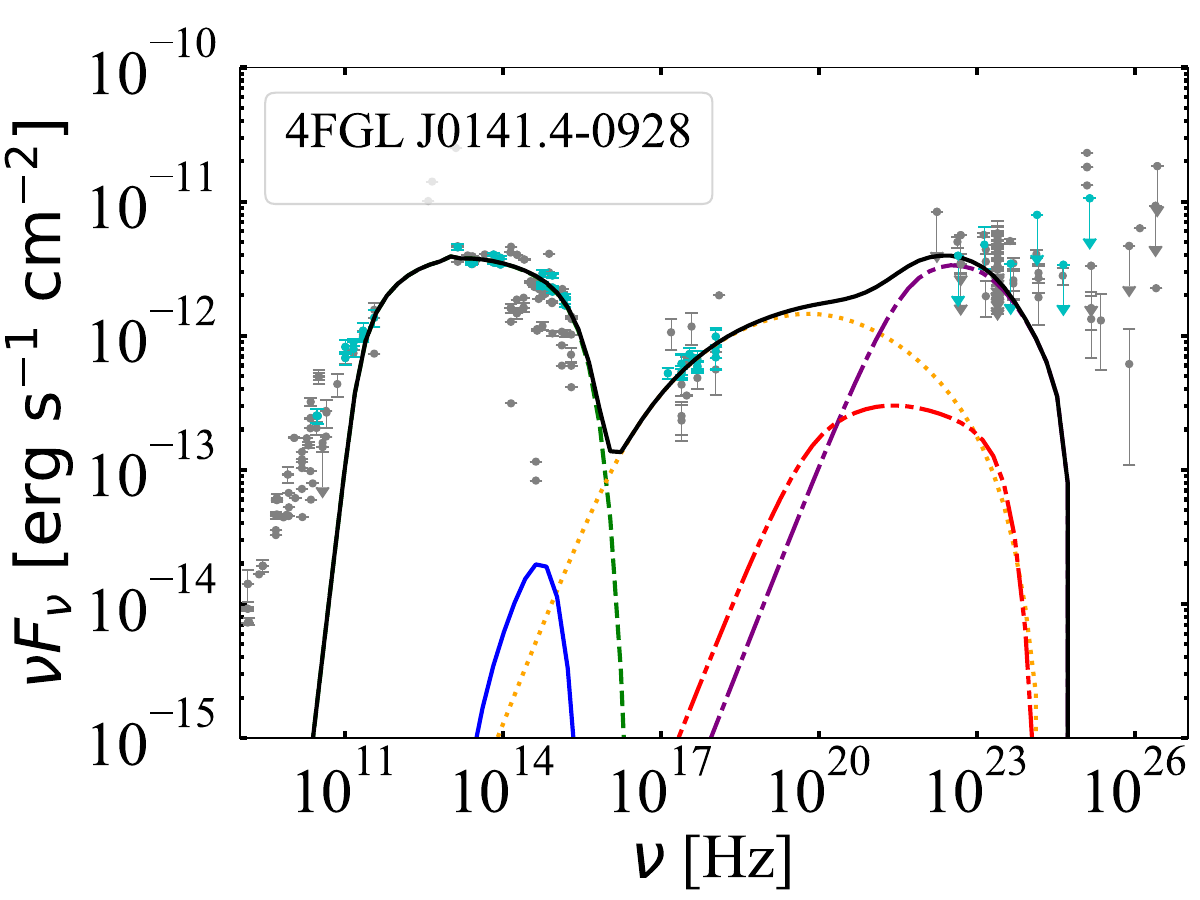}

\includegraphics[width=8cm,height=5cm]{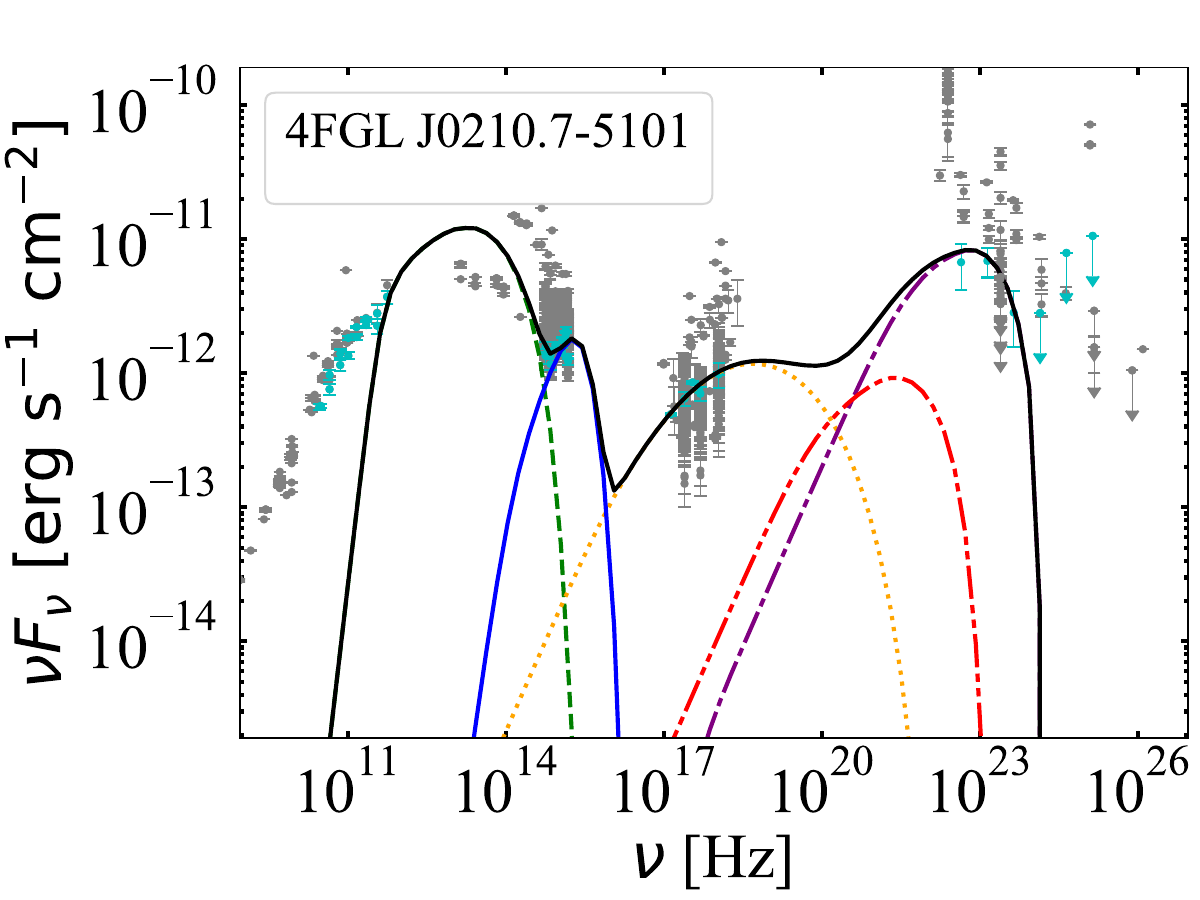}
\hspace{1.2cm}
\includegraphics[width=8cm,height=5cm]{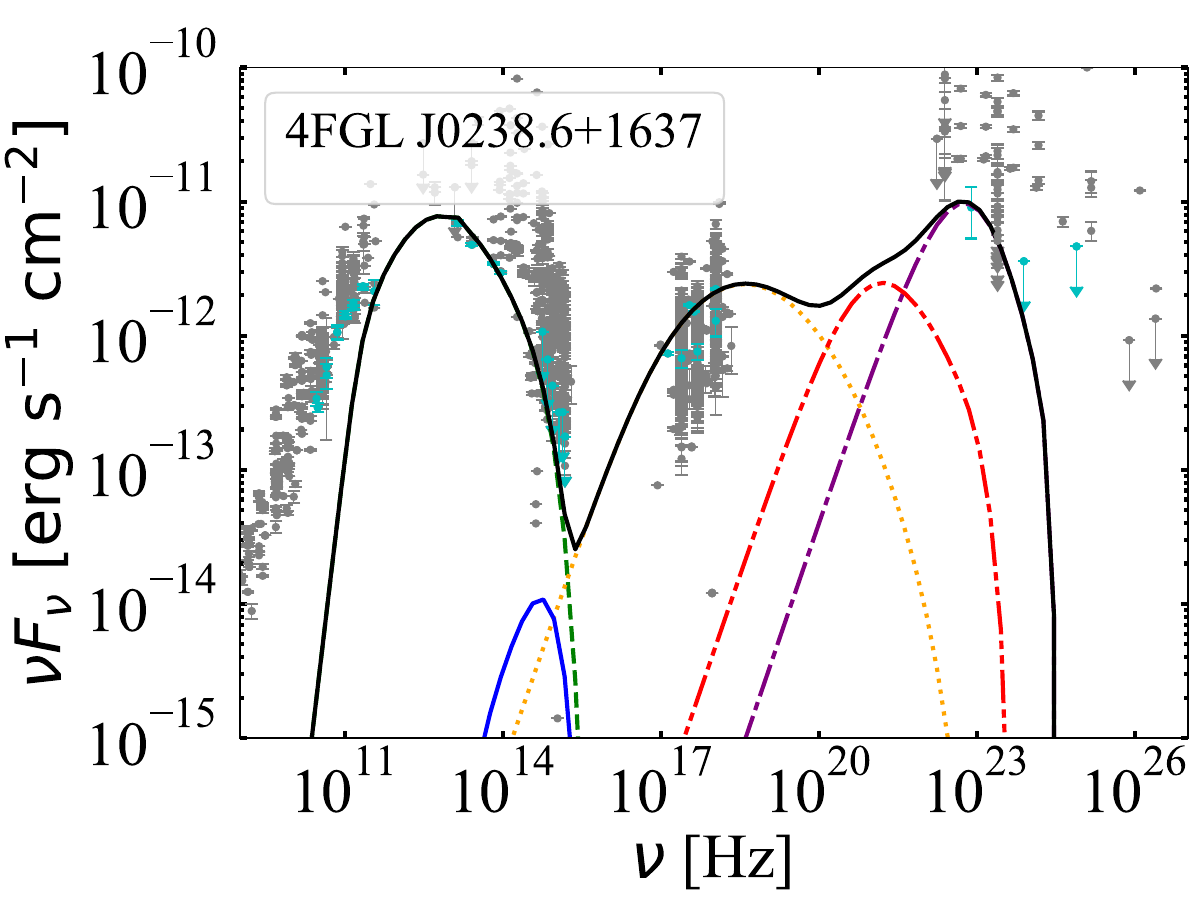}

\includegraphics[width=8cm,height=5cm]{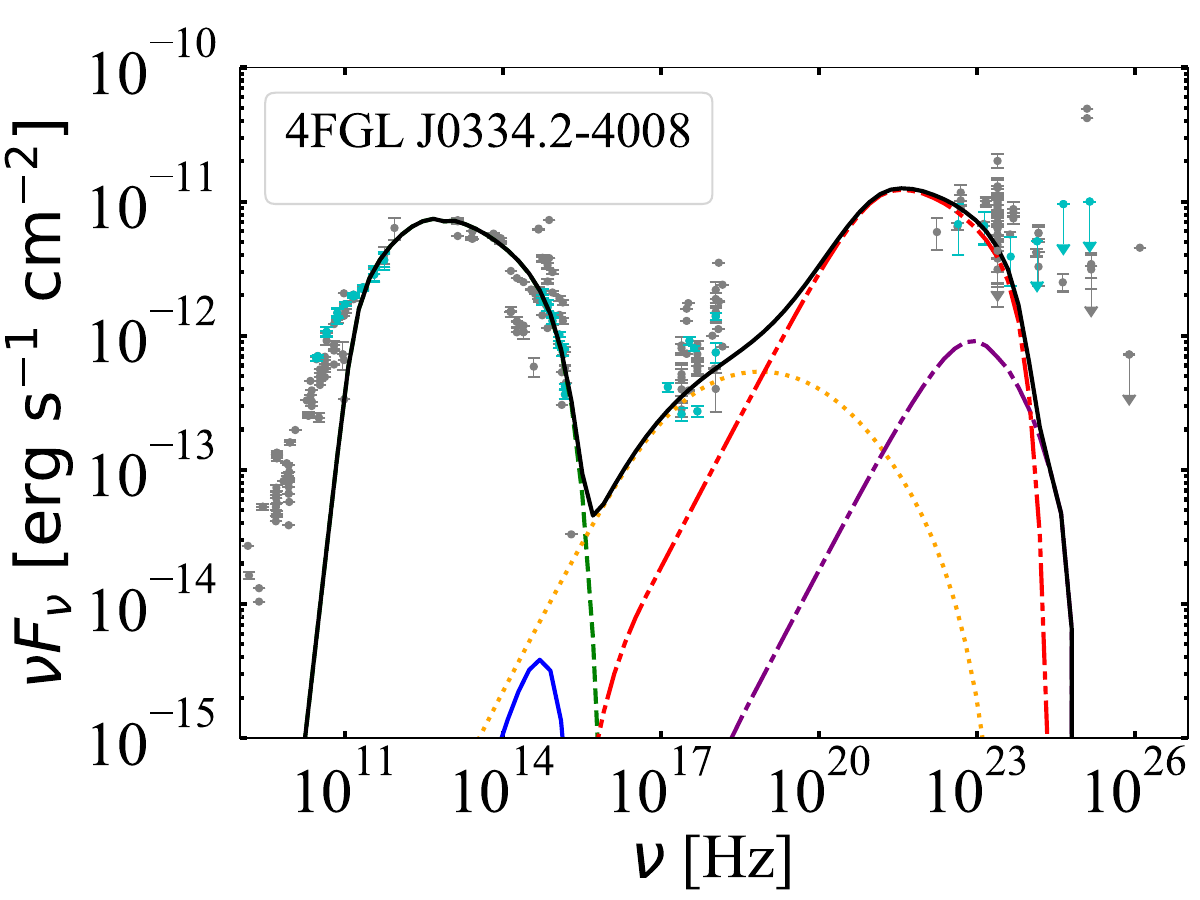}
\hspace{1.2cm}
\includegraphics[width=8cm,height=5cm]{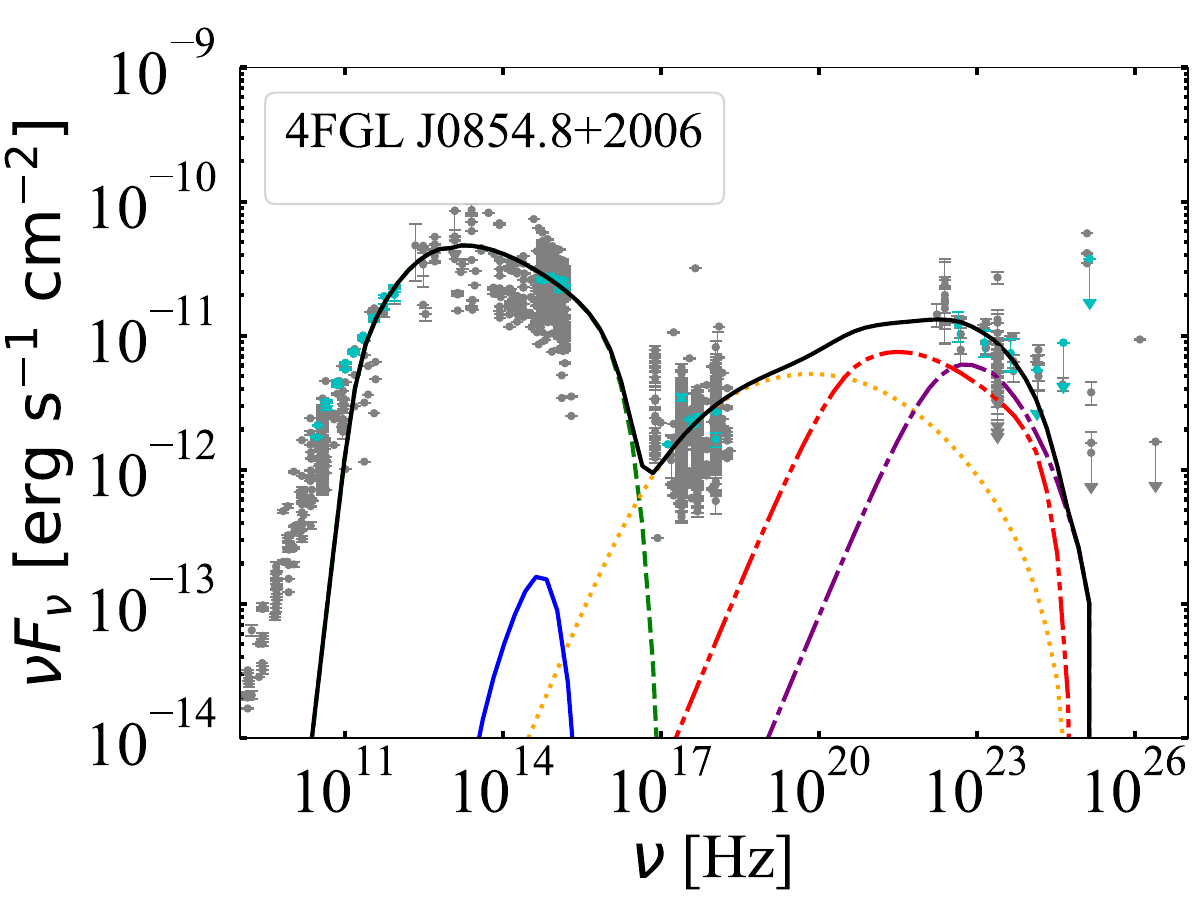}

\includegraphics[width=8cm,height=5cm]{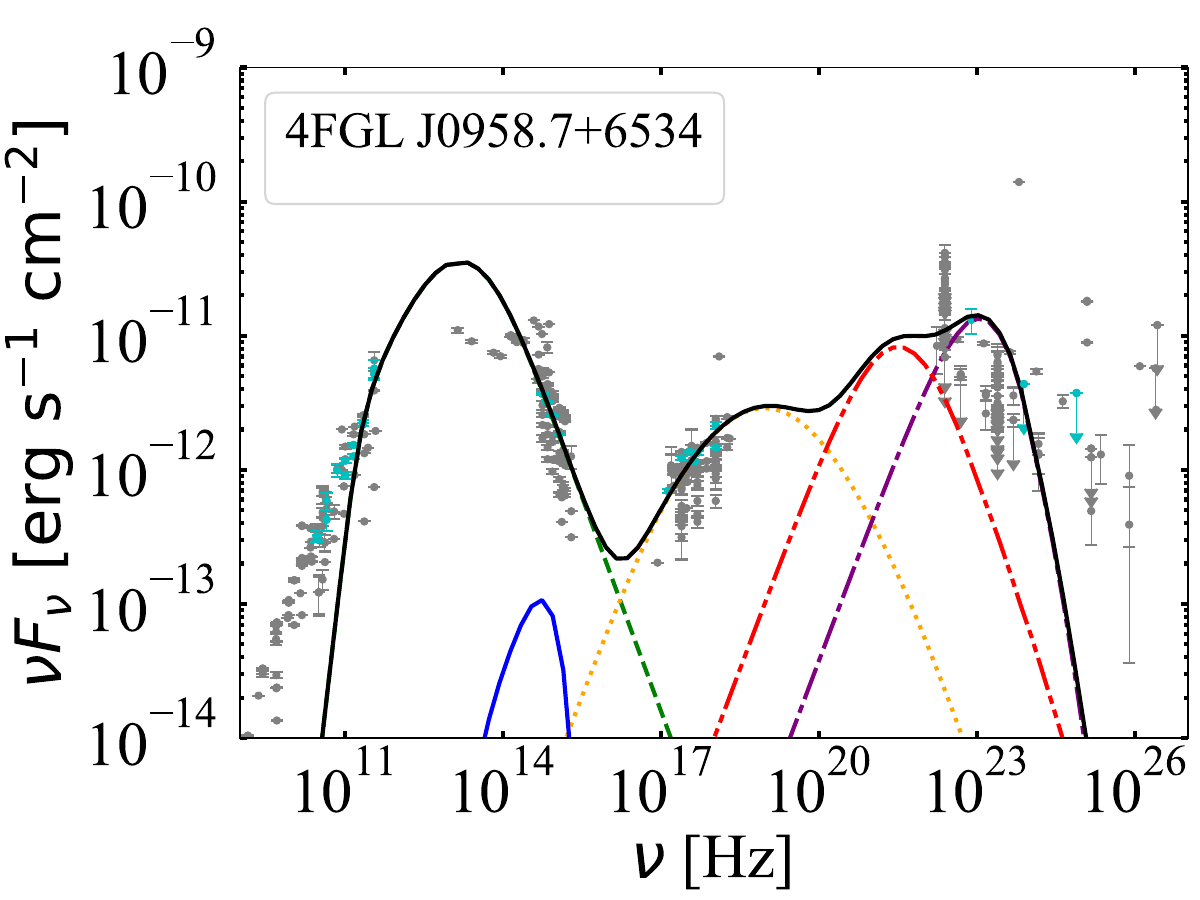}
\hspace{1.2cm}
\includegraphics[width=8cm,height=5cm]{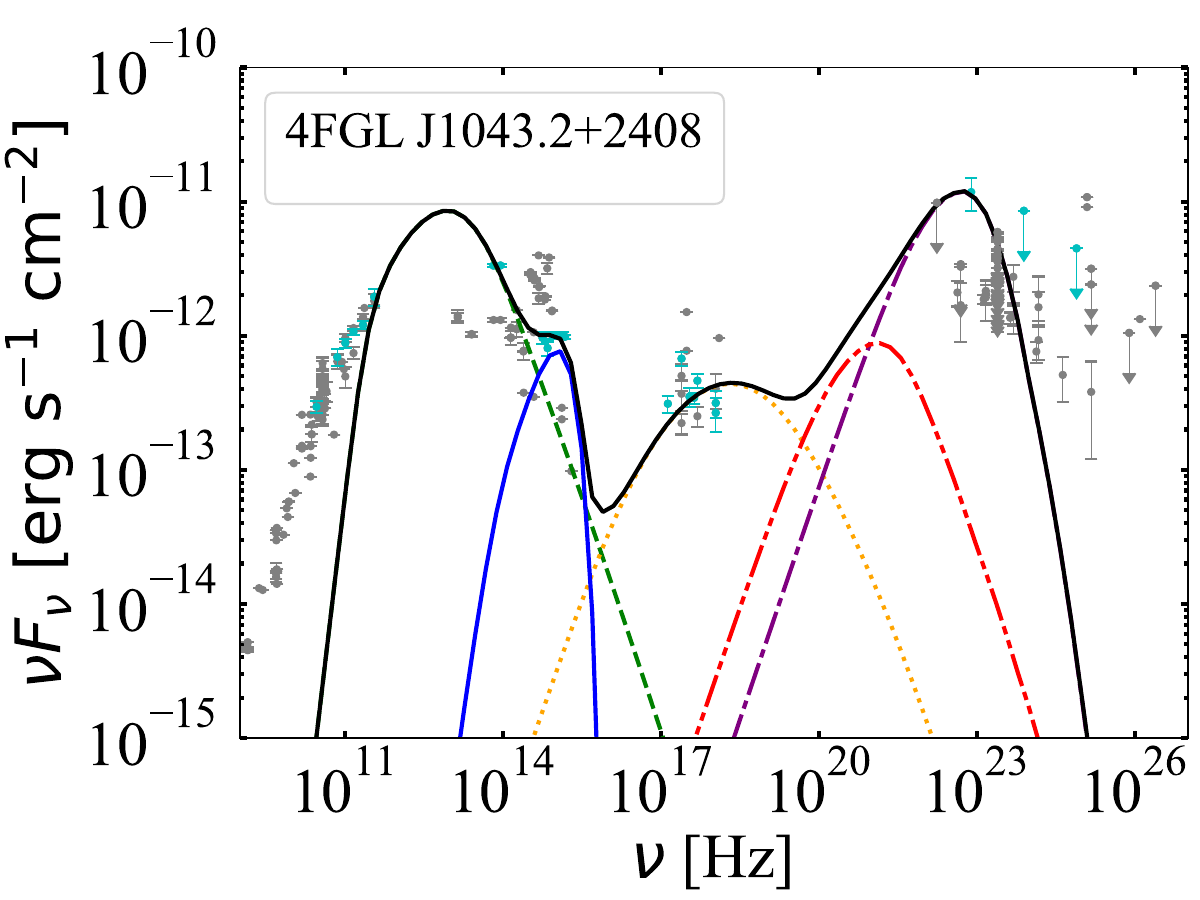}

\caption{The fitting results of the SEDs of 14 LBLs with the conventional
one-zone EC model. The gray points are archival data from SSDC,
and the cyan points from infrared to $\gamma$-ray bands are quasi-simultaneous data.
The green dashed line represents the synchrotron emission,
the yellow dotted line represents the SSC emission,
the purple and red dash-dotted lines represent the EC emission,
in which soft photons are from the BLR and DT, respectively,
the blue solid line in the optical-UV band represents the thermal radiation of the accretion disk,
and the black solid curve is the total emission obtained
by summing all the emission components.
It should be noted that, in the framework of the one-zone leptonic
model, the synchrotron emission below the turnover frequency
(typically $\nu\mathrm{^{obs}}<10^{11}~\mathrm{Hz}$) is inevitably self-absorbed.
Therefore, we do not explain the corresponding radio data throughout this work.
}
 
\end{figure}
\addtocounter{figure}{-1}
\begin{figure}
\centering
\includegraphics[width=8cm,height=5cm]{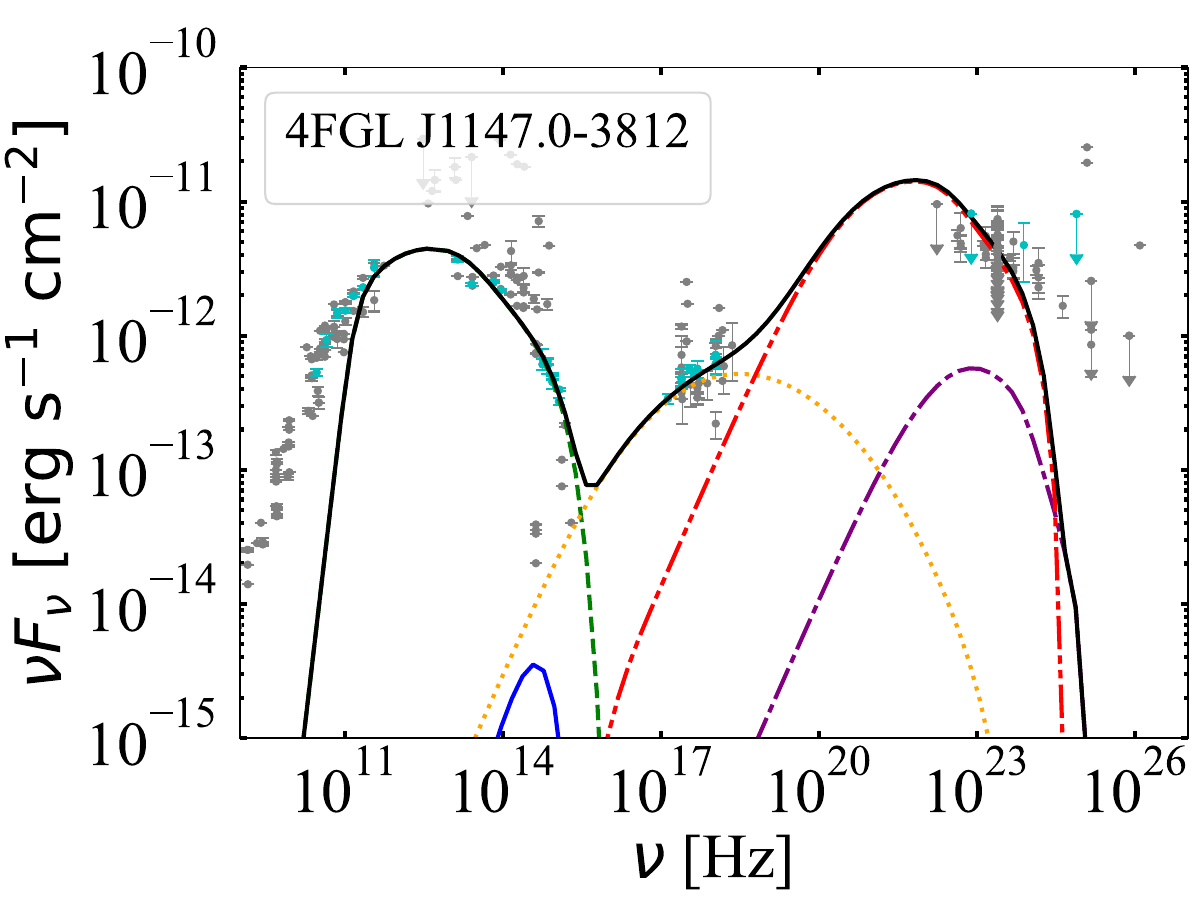}
\hspace{1.2cm}
\includegraphics[width=8cm,height=5cm]{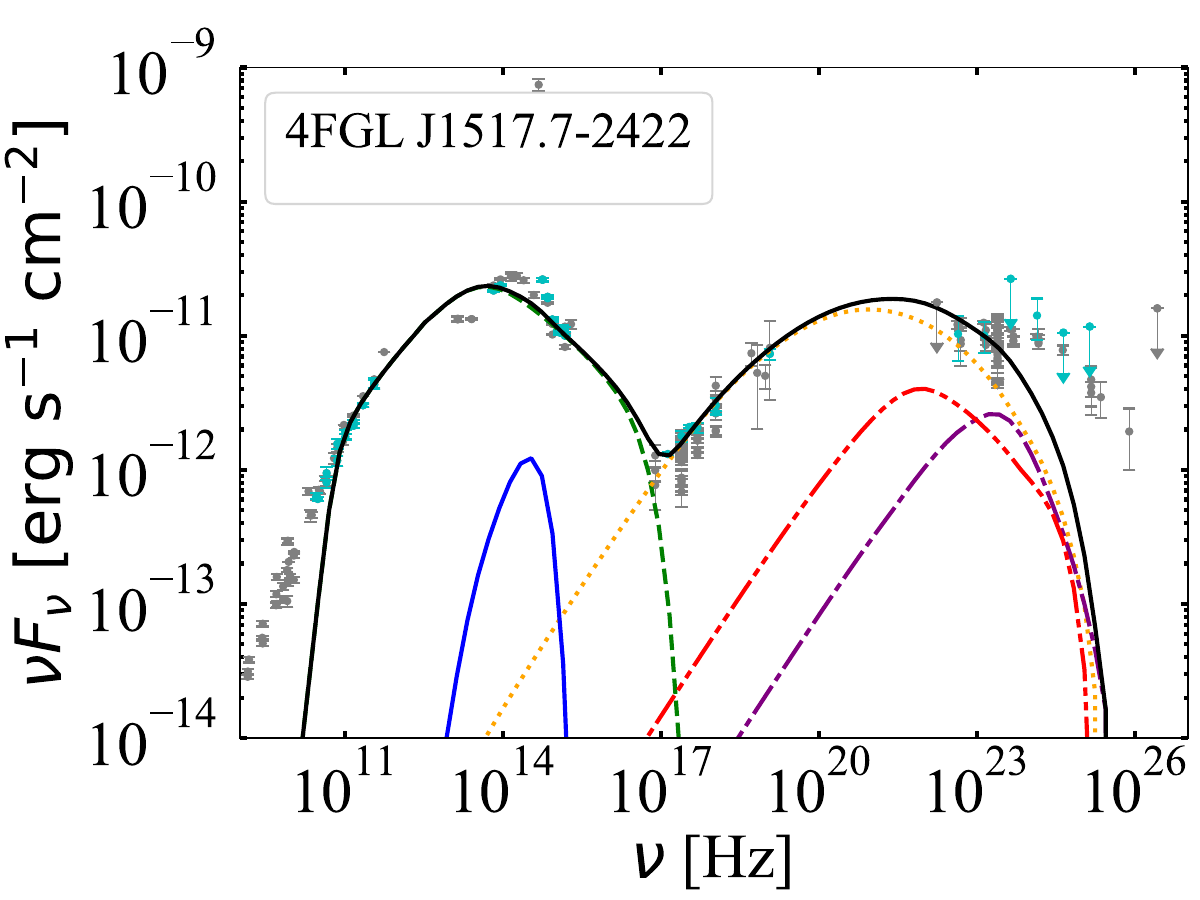}

\includegraphics[width=8cm,height=5cm]{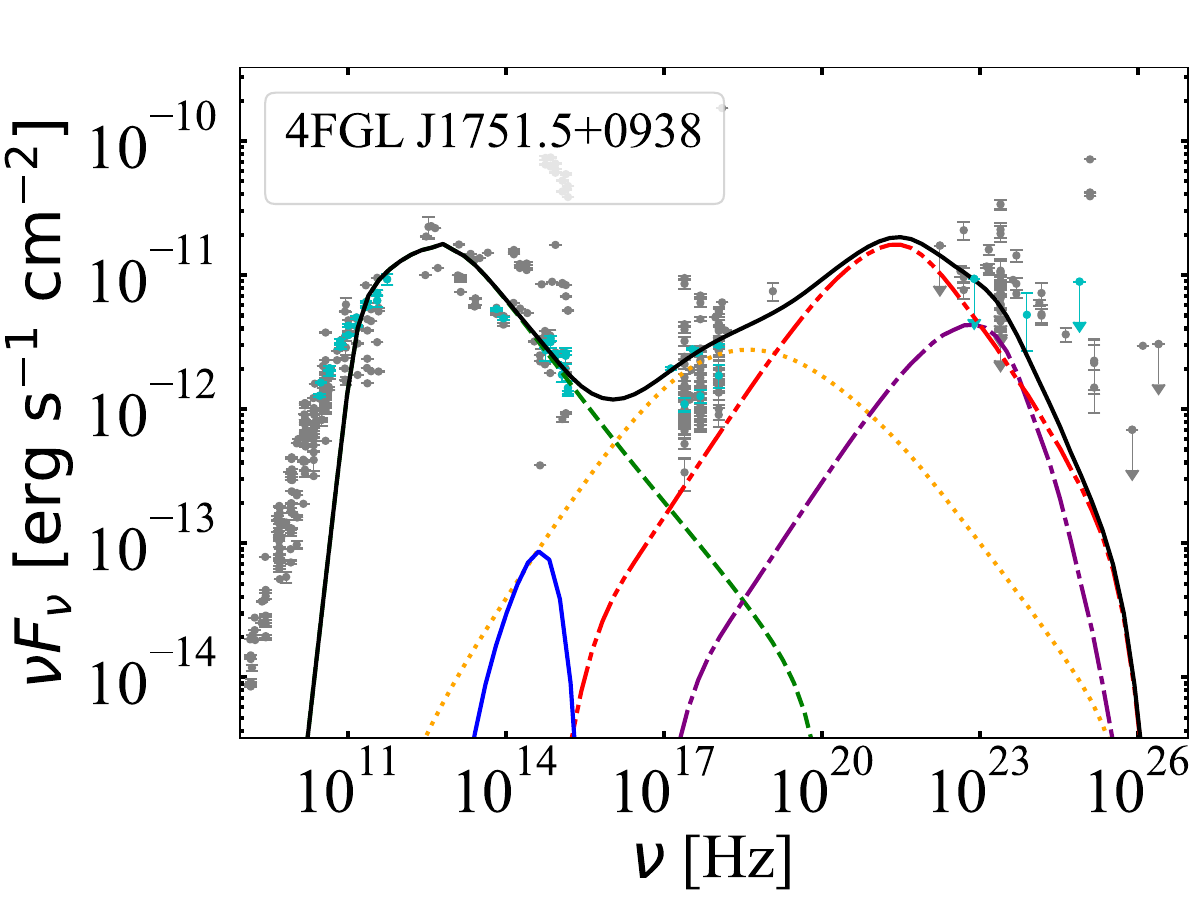}
\hspace{1.2cm}
\includegraphics[width=8cm,height=5cm]{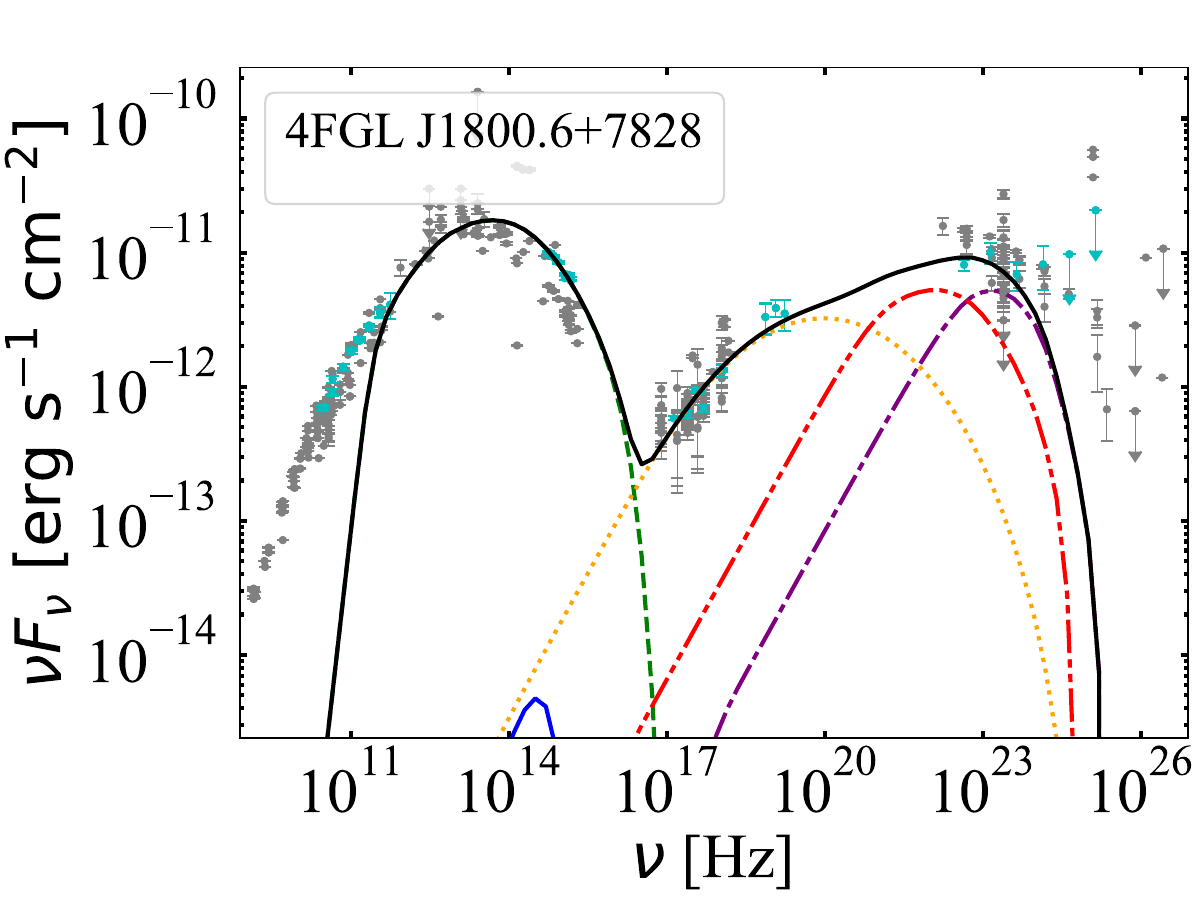}

\includegraphics[width=8cm,height=5cm]{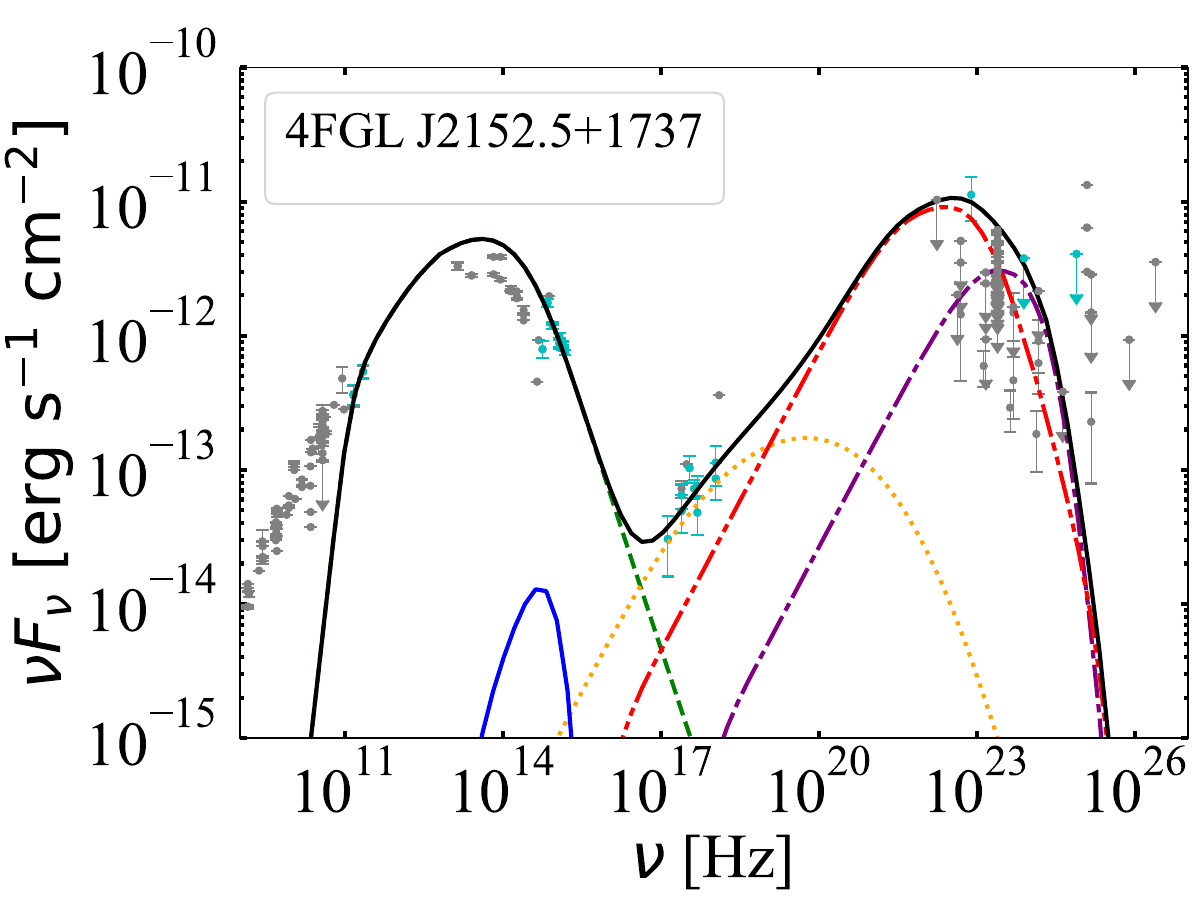}
\hspace{1.2cm}
\includegraphics[width=8cm,height=5cm]{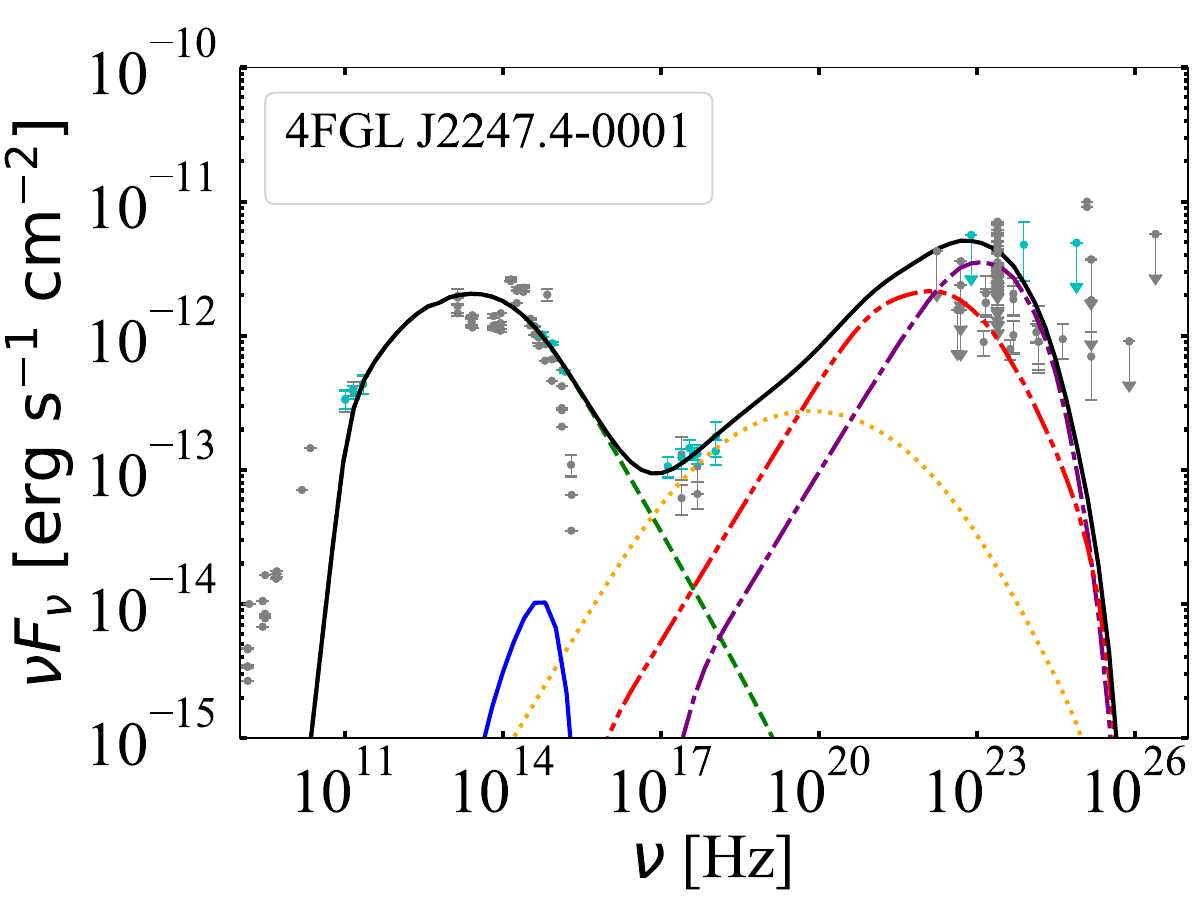}

\caption{-$continued$.}
\label{fig2}
\end{figure}
\twocolumn

\begin{figure}
\centering

\includegraphics[width=1\columnwidth]{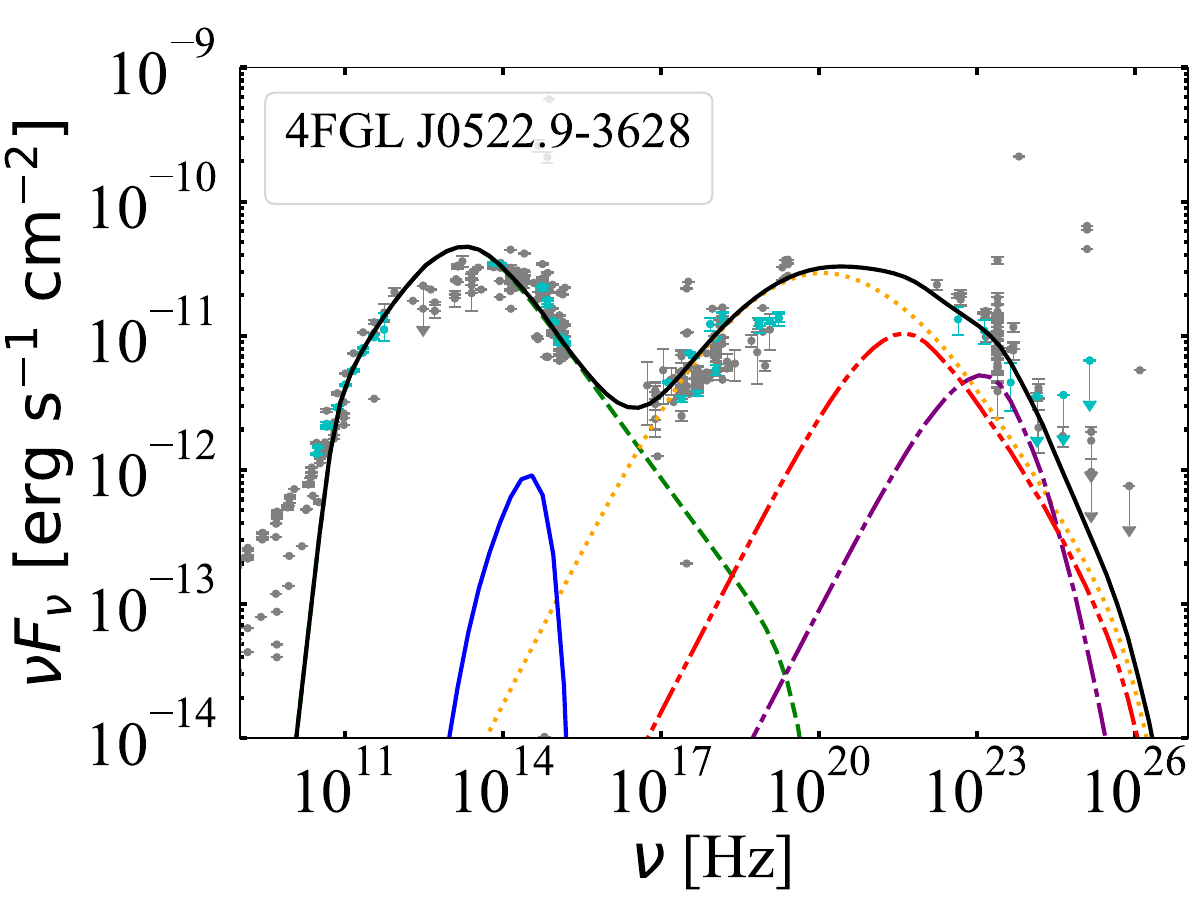}
\includegraphics[width=1.1\columnwidth]{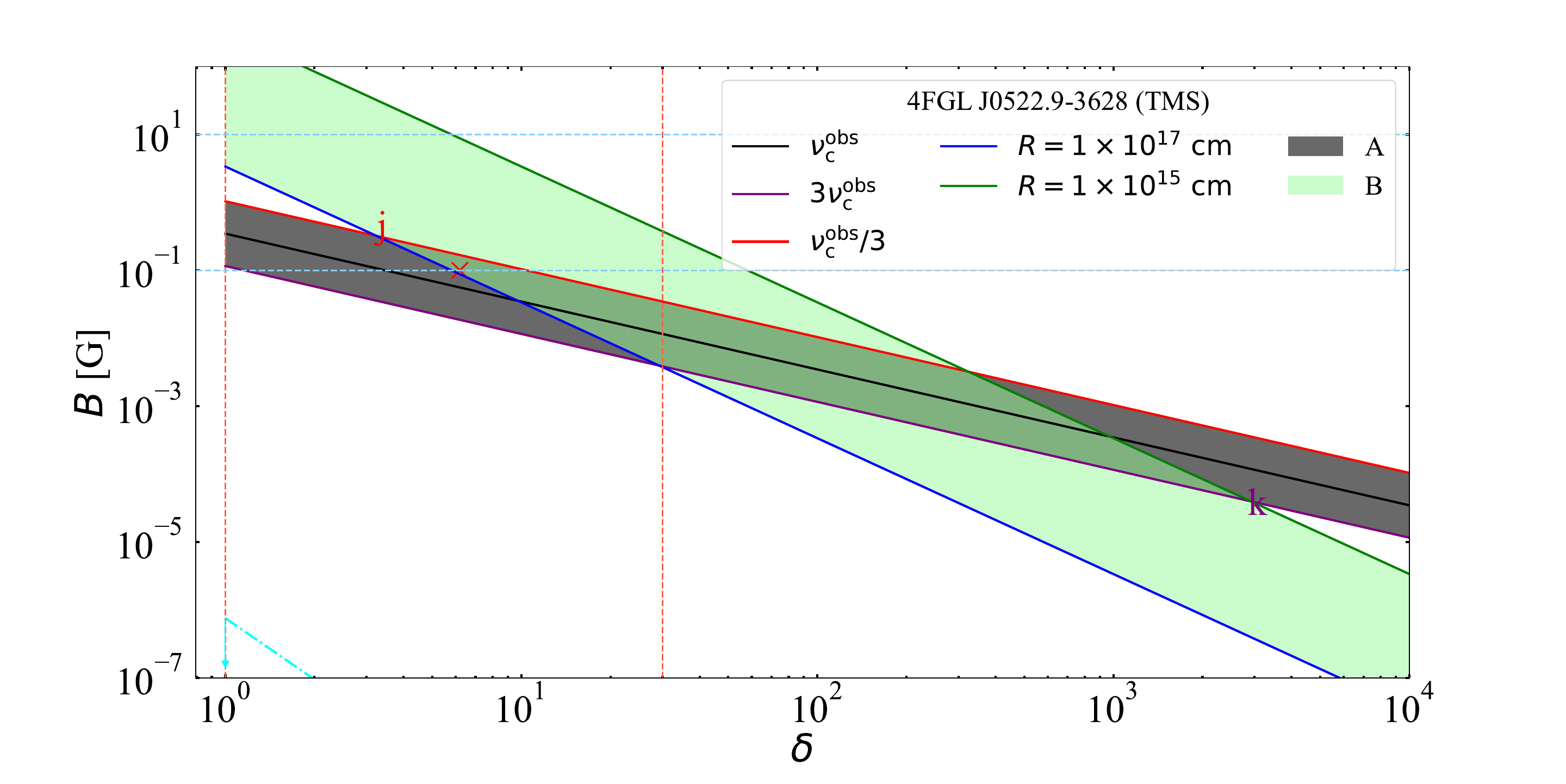}
\includegraphics[width=1\columnwidth]{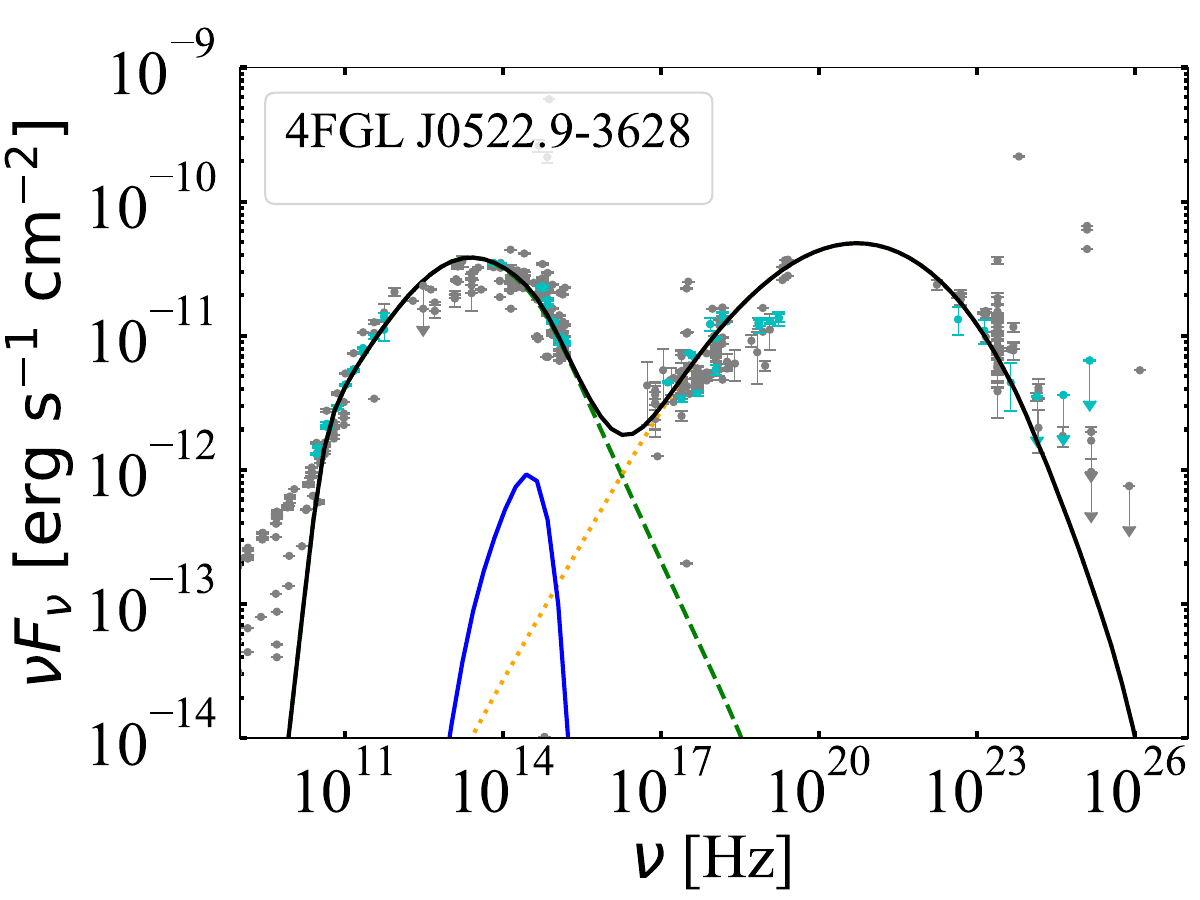}

\caption{Upper panel: the fitting result of the SED of 4FGL J0522.9-3628
with the one-zone EC model. The symbols and lines have
the same meaning as in Fig.~\ref{fig2}.
Middle panel: the parameter space of the one-zone SSC model in the TMS regime.
Symbols j and k represent the upper left and lower right points of
the intersection area of A and B, respectively.
Other symbols and lines are the same as shown in Fig.~\ref{fig1}.
Lower panel: the same as the upper panel, but using the one-zone SSC model
(the adopted $\delta$ and $B$ correspond to the red cross in the middle panel).}
\label{fig3}
\end{figure}

\begin{figure}
\centering

\includegraphics[width=1.2\columnwidth]{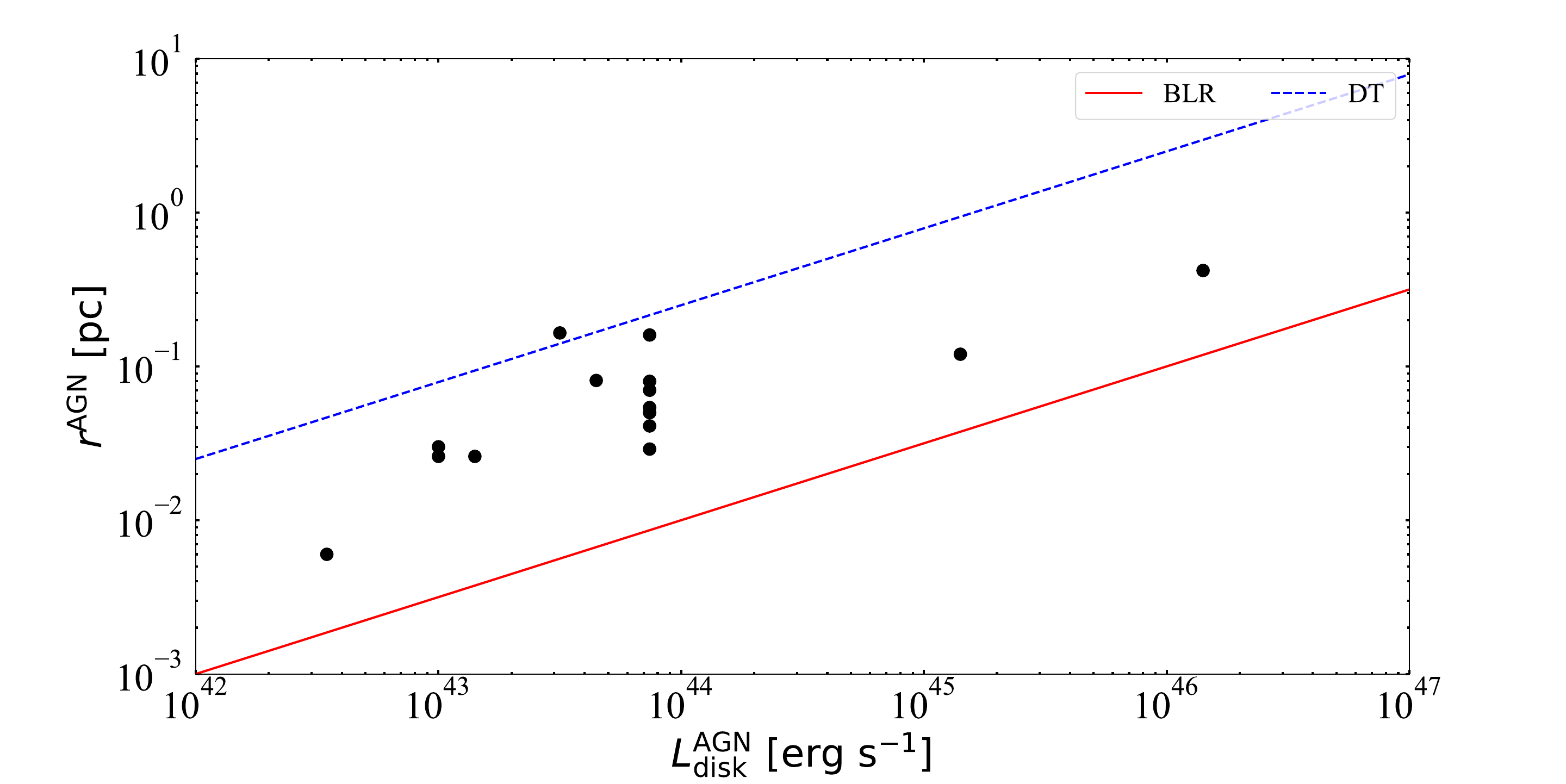}
\includegraphics[width=1.2\columnwidth]{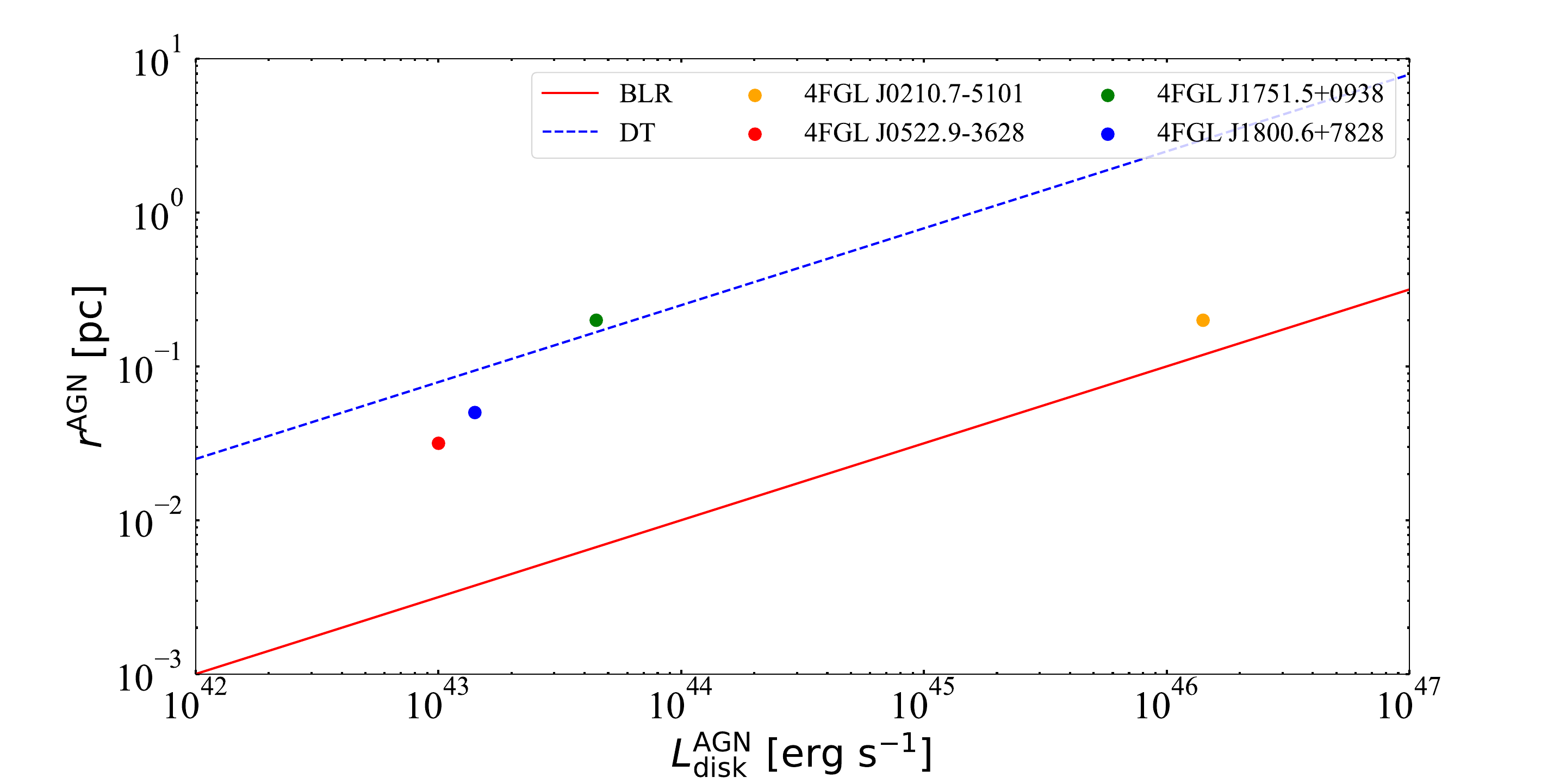}
\includegraphics[width=1.2\columnwidth]{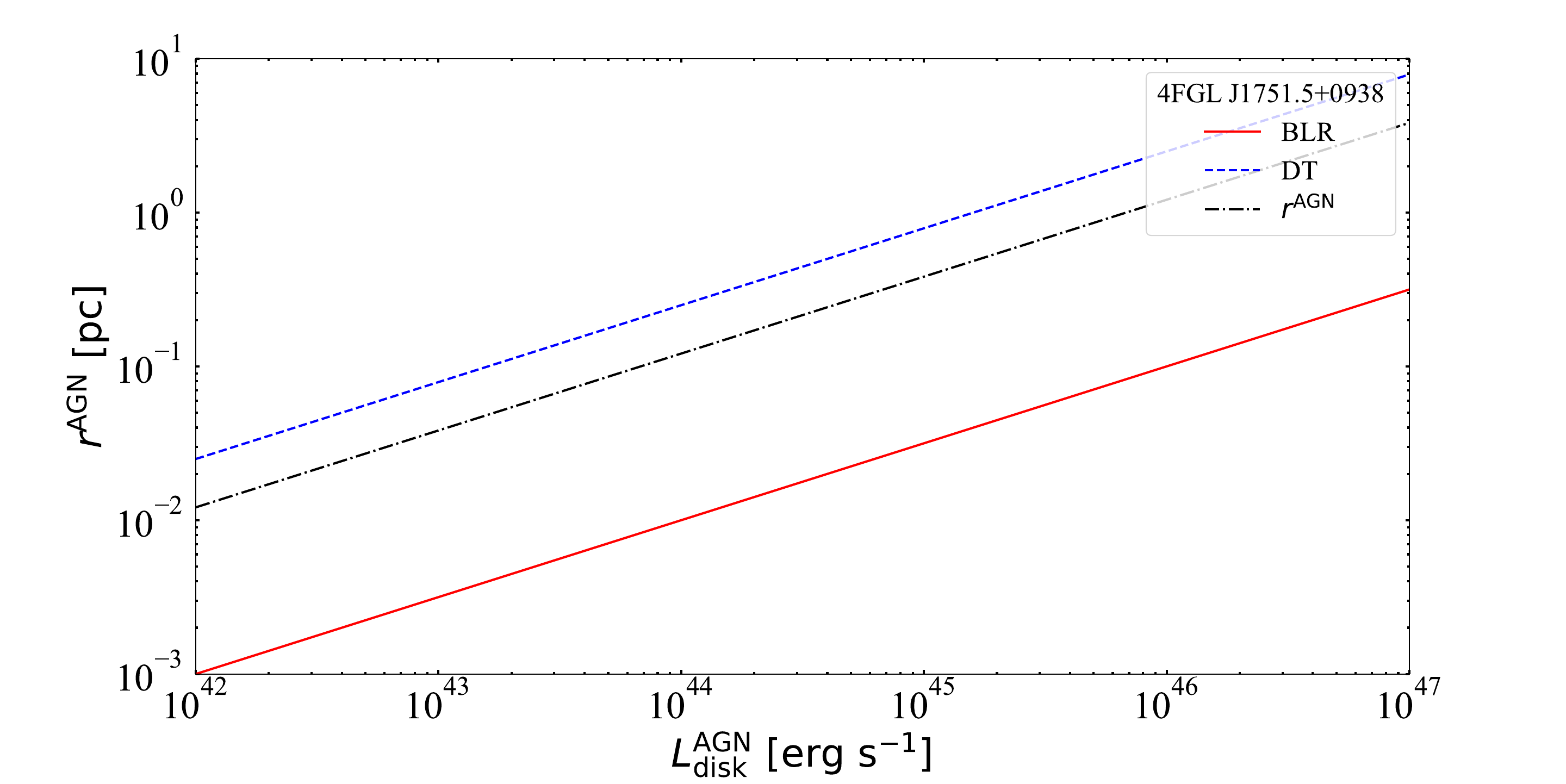}

\caption{The distance between the $\gamma$-ray emitting region and the SMBH
($r\mathrm{^{AGN}}$) as a function of the luminosity of an accretion disk
($L\mathrm{_{disk}^{AGN}}$). The upper panel and the middle panel are obtained by
SEDs fitting and variability timescale calculation, respectively.
The lower panel shows the relation between $r\mathrm{^{AGN}}$ and
$L\mathrm{^{AGN}_{disk}}$ for 4FGL J1751.5+0938, which is obtained when
$U\mathrm{_{ext}}$ remains unchanged.
The sample sources are shown in solid points. The red solid line
and the blue dashed line represent the characteristic distances of the BLR and DT,
respectively. The meaning of other symbols and lines are given in the legends.}
\label{fig4} 
\end{figure}

\begin{figure}
\centering

\includegraphics[width=1\columnwidth]{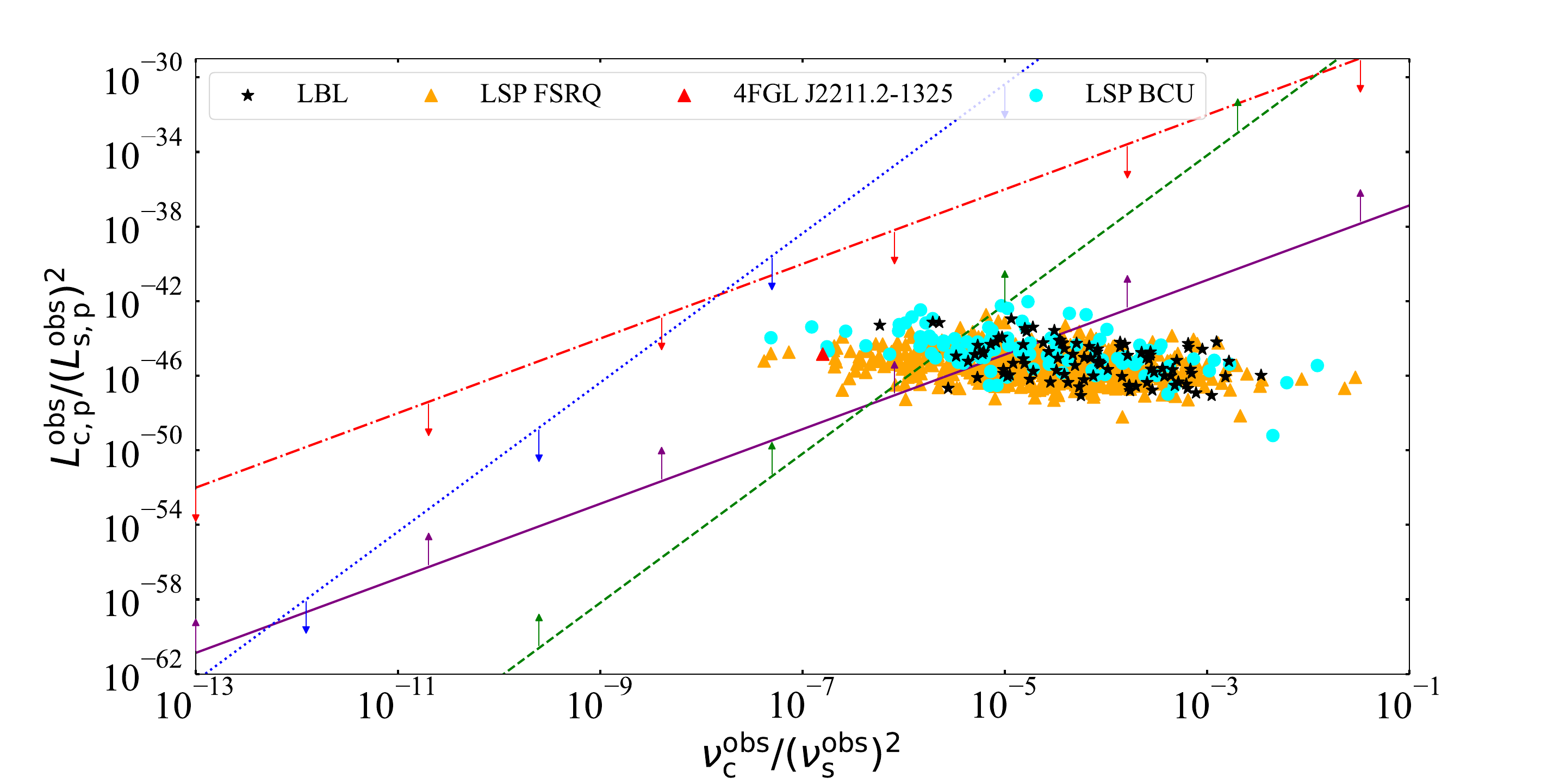}

\caption{The distribution of all types of LSP blazars in $Fermi$-4LAC
(using archival data) in the constraints of $\delta\mathrm{_{j}}\le30$
(the purple solid line with arrows), $B\mathrm{_{j}}\ge0.1~\mathrm{G}$
(the green dashed line with arrows), $\delta\mathrm{_{k}}\ge1$
(the red dash-dotted line with arrows), $B\mathrm{_{k}}\le10~\mathrm{G}$
(the blue dotted line with arrows).
The meaning of other symbols are given in the legends.}
\label{fig5}
\end{figure}

\begin{figure}
\centering

\includegraphics[width=1.1\columnwidth]{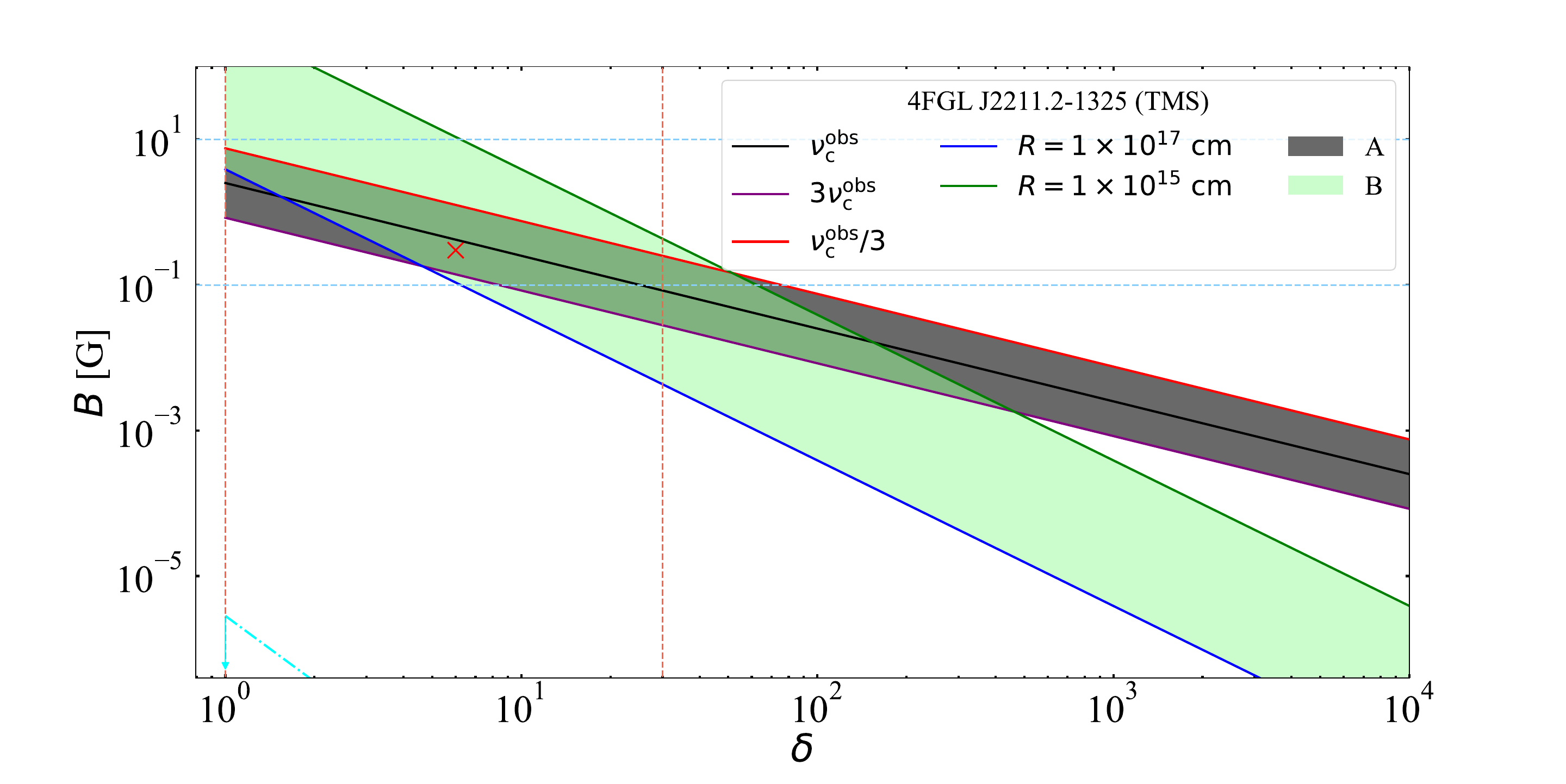}
\includegraphics[width=1\columnwidth]{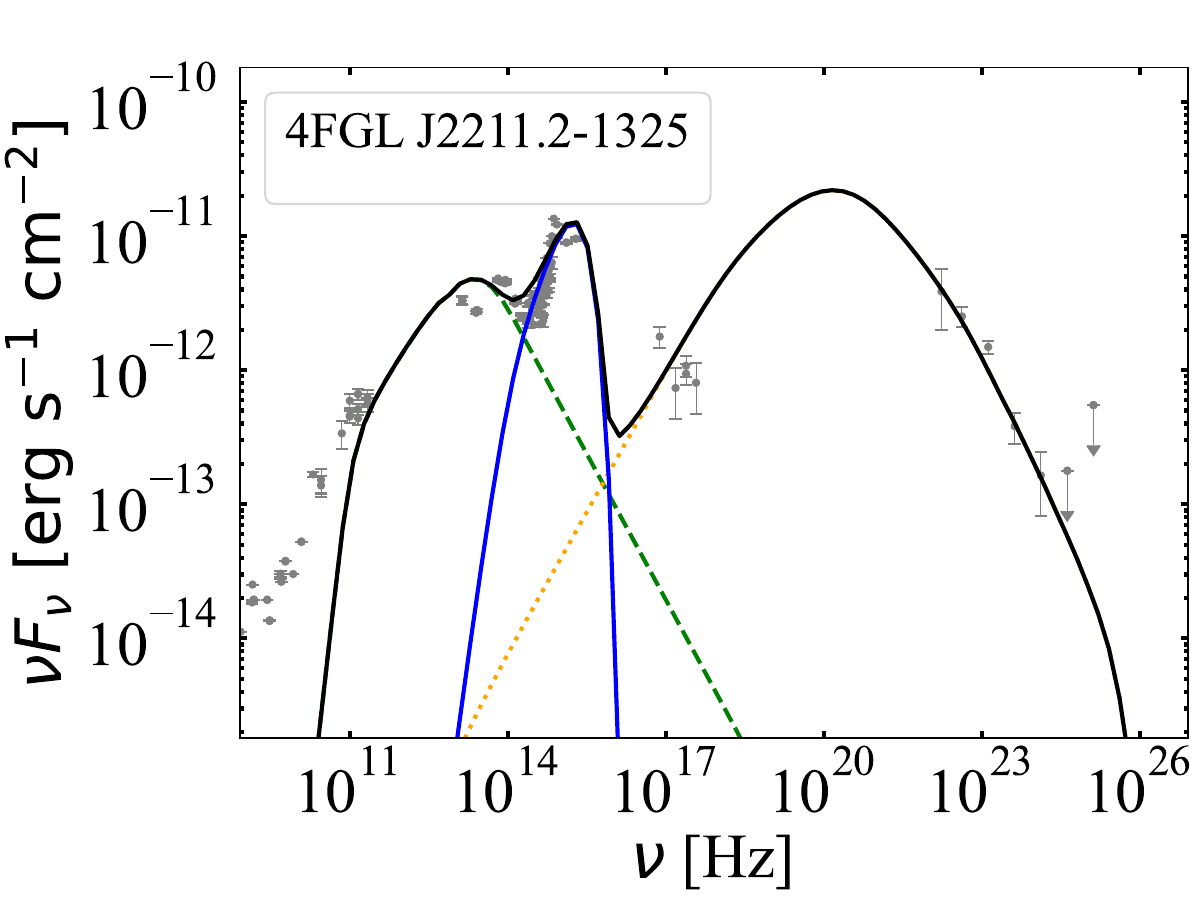}

\caption{Upper panel: the parameter space of 4FGL J2211.2-1325 in the TMS regime. The symbols and lines are the same as shown in Fig.~\ref{fig1}.
Lower panel: the fitting result of the SED with the one-zone SSC model
(the adopted $\delta$ and $B$ correspond to the red cross in the upper panel).
The symbols and lines have
the same meaning as in Fig.~\ref{fig2}.}
\label{fig6}
\end{figure}

\subsection{The physical properties of LSPs}
As we can see from the middle panel of Fig.~\ref{fig3} that at least four conditions,
i.e., $\delta\mathrm{_{j}}\le30$, $B\mathrm{_{j}}\ge0.1~\mathrm{G}$,
$\delta\mathrm{_{k}}\ge1$, $B\mathrm{_{k}}\le10~\mathrm{G}$,
are required to find the parameter space under the observational constraints.
Combining Eq.~(\ref{eq2}) and Eq.~(\ref{eq5}), $\delta$ and $B$ can be expressed as
\begin{equation}\label{eq22}
\delta=\frac{2.8\times10^{6}}{1+z}\left[\frac{2f(\alpha_{1},\alpha_{2})}{cR^{2}}\right]^{1/2}\frac{\nu\mathrm{_{c}^{obs}}}{(\nu\mathrm{_{s}^{obs}})^{2}}\left[\frac{L\mathrm{_{c,p}^{obs}}}{(L\mathrm{_{s,p}^{obs}})^{2}}\right]^{-1/2}
\end{equation}
and
\begin{equation}\label{eq23}
B=\left(\frac{1+z}{2.8\times10^{6}}\right)^{2}\left[\frac{cR^{2}}{2f(\alpha_{1},\alpha_{2})}\right]^{1/2}\left[\frac{\nu\mathrm{_{c}^{obs}}}{(\nu\mathrm{_{s}^{obs}})^{2}}\right]^{-2}\left[\frac{L\mathrm{_{c,p}^{obs}}}{(L\mathrm{_{s,p}^{obs}})^{2}}\right]^{1/2},
\end{equation}
respectively.
Here we set $f(\alpha_{1},\alpha_{2})=3$, because it has little effect on the results.
Then, the sources and the above conditions can be represented as points and regions in the
$\nu\mathrm{_{c}^{obs}}/(\nu\mathrm{_{s}^{obs}})^{2}-L\mathrm{_{c,p}^{obs}}/(L\mathrm{_{s,p}^{obs})^{2}}$ diagram, respectively.
Since the analytical method is independent of the specific subclass of blazar,
we can apply it to all types of LSPs in $Fermi$-4LAC
to explore their intrinsic physical properties.
To apply the above method, the values of peak frequency and peak luminosity are needed.
Fortunately, \citet{2022ApJS..262...18Y, 2023SCPMA..6649511Y}
have constructed the historical SEDs of all 750 FSRQs and 844 BL Lacs with measured
redshift in the Fourth $Fermi$-LAT 12-year Source catalog \citep{2022ApJS..260...53A},
and fitted them with a parabolic equation, enabling the derivation of the peak frequency and peak luminosity.
Taking advantage of the archival data, we show the above constraints
and the distribution of LSPs in Fig.~\ref{fig5}.
It can be seen that most of the LBLs cannot satisfy these constraints simultaneously.
This implies that the external photon field is more important for LBLs than for
IBLs and HBLs, which is consistent with the conclusion of
\citet{2014ApJ...788..104Z}.
Therefore, our results suggest that LBLs should be masquerading BL Lacs.
On the other hand, since the IC process of the external photon field will significantly
increase the total luminosity of the blazar and the ratio of
$L\mathrm{^{obs}_{c}}/L\mathrm{^{obs}_{s}}$
\citep[e.g.,][]{2012ApJ...752..157Z},
we suggest that the LBLs should be more Compton dominated than HBLs
\citep[e.g.,][]{2014MNRAS.439.2933Y},
and have higher luminosity, which is consistent with the blazar sequence
\citep[e.g.,][]{1998MNRAS.299..433F, 2011ApJ...740...98M, 2017MNRAS.469..255G}.

Interestingly, Fig.~\ref{fig5} shows that the high-energy peak of some FSRQs
can be explained by the SSC emission.
In order to illustrate this fact better, we take 4FGL J2211.2-1325 as an example
(randomly selected from the sample), we show its parameter space under the TMS regime
and fitting results with the one-zone SSC model in the upper and lower panels of
Fig.~\ref{fig6}, respectively.
It can be seen that the importance of the external photon field for FSRQs can be low,
which indicates that the emitting region may be located far from the SMBH,
and this can be tested by the variability timescale as mentioned before.

Since the same population of relativistic electrons produce the SSC and EC emissions
in the one-zone EC model, we have $\nu\mathrm{_{c}^{obs}}\approx\nu\mathrm{_{EC}^{obs}}=(4/3)\gamma\mathrm{_{b}^{2}}\nu\mathrm{_{soft}^{AGN}}\Gamma\delta/(1+z)$,
where $\nu\mathrm{^{obs}_{EC}}$ is the peak frequency of the EC emission.
Combining Eq.~(\ref{eq1}), we can obtain $\nu\mathrm{_{c}^{obs}}/(\nu\mathrm{_{s}^{obs}})^{2}\propto (1+z)\nu\mathrm{_{soft}^{AGN}}/(\gamma\mathrm{_{b}^{2}}B^{2})$.
On the other hand, for the case of the one-zone SSC model,
$\nu\mathrm{_{c}^{obs}}/(\nu\mathrm{_{s}^{obs}})^{2}$ can be expressed as
$\nu\mathrm{_{c}^{obs}}/(\nu\mathrm{_{s}^{obs}})^{2}\propto (1+z)/(\delta B)$,
according to Eq.~(\ref{eq2}).
Since $z$ has little effect on the results, and LBLs and FSRQs have the same
$\nu\mathrm{_{soft}^{AGN}}$, the difference in the distribution of LBLs and FSRQs
shown in Fig.~\ref{fig5} indicates that when the one-zone SSC model is applicable,
FSRQs have larger $\delta B$, otherwise FSRQs have smaller $\gamma\mathrm{_{b}}B$.
In addition, we can also find in Fig.~\ref{fig5} that the distribution of
blazar candidates of unknown types (BCUs) is similar to that of FSRQs.

\begin{figure}
\centering

\includegraphics[width=1\columnwidth]{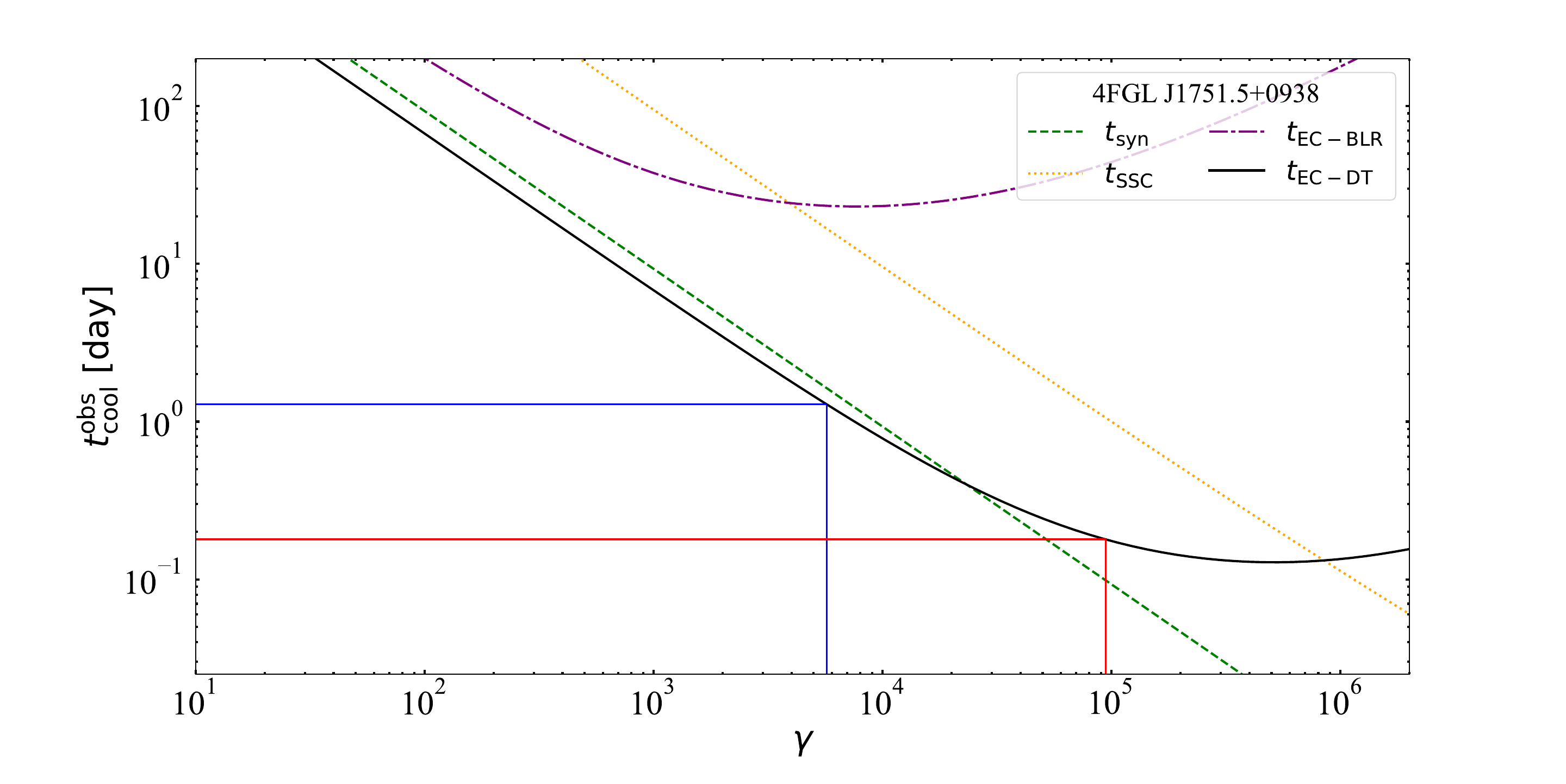}

\caption{The radiative cooling timescales of relativistic electrons
$t\mathrm{_{cool}^{obs}}$ as a function of the electron Lorentz factors $\gamma$
for 4FGL J1751.5+0938.
The blue and red solid curves represent the Lorentz factors and the corresponding
EC-DT cooling timescales of the electrons that produce 500 MeV and 150 GeV radiations,
respectively. The meaning of other line styles are given in the legends.}
\label{fig7}
\end{figure}

\subsection{Variability timescale and the determination of high-energy peak origin}
Normally, the variability timescale
\citep[e.g.,][]{2010ApJ...712..957A, 2010ApJ...715..362J, 2010MNRAS.405L..94T, 2011MNRAS.418...90L, 2012JPhCS.355a2032A, 2012ApJ...751L...3G, 2013MNRAS.431..824B, 2015MNRAS.452.1280R}
may be a useful indicator to the location of emitting region and the origin of
the high-energy peak. When the BLR and DT are taken into account, i.e.,
$U\mathrm{_{ext}}\ne0$, in order to explain the high-energy peak with the SSC emission,
$U\mathrm{_{syn}}>U\mathrm{_{ext}}$ must be satisfied.
Here, we take 4FGL J0522.9-3628 as an example to explore the conditions for
the applicability of the one-zone SSC model.
For simplicity, we assume that the IC scattering occurs in the TMS regime, then
$R>5.4\times10^{15}~\mathrm{cm}$ is required when $U\mathrm{_{syn}}>U\mathrm{_{BLR}}$,
and $R>5.7\times10^{16}~\mathrm{cm}$ for the case of $U\mathrm{_{syn}}>U\mathrm{_{DT}}$.
Then, we find that the SSC emission can only dominate the high-energy peak when
$t\mathrm{_{var}^{obs}}\approx R(1+z)/(c\delta)>2.3~\mathrm{day}$.
In that case, the contribution of the EC emission to the spectrum is negligible.
Therefore, our results suggest that the one-zone SSC model is only possible to be used to
interpret the observed SEDs in the case of relatively long variability timescale.
It should be noted that the reference range of the variability
timescale mainly depends on the range of $R$.
Since the value of $R$ is mainly affected by the ratio of
$L\mathrm{^{obs}_{c}}/L\mathrm{^{obs}_{s}}$,
which can vary greatly among different sources, there may be large differences
between the thresholds of the variability timescale for different sources.
Therefore, it is possible that the source with
the shortest variability timescale, such as 4FGL J0522.9-3628 in the middle panel of
Fig.~\ref{fig4}, can be fitted with a one-zone SSC model, while the sources with
longer variability timescales can only be fitted with a one-zone EC model.
However, the variability timescale of 4FGL J0522.9-3628 used to estimate
the location of the $\gamma$-ray emitting region is only 0.43 days.
It seems to be inconsistent with our conclusion.
However, it should be noted that \citet{2013ApJ...767..103V} only presented
the minimum $\gamma$-ray variability timescales, which may correspond to different SEDs
from those we collected.
Since the analytical results depend on the peak frequency and peak luminosity,
the parameter space that satisfies the observational constraints in the TMS regime may
not be found anymore.
More specifically, for 4FGL J0522.9-3628, we believe that the high-energy peak should
originate from the EC emission due to the short variability timescale, resulting in
a larger $\nu\mathrm{_{c}^{obs}}/(\nu\mathrm{_{s}^{obs}})^{2}$, thus locating outside the
region that can satisfy four conditions simultaneously (see Fig.~\ref{fig5}).
Therefore, its SED may cannot be fitted by the one-zone SSC model anymore.
The analysis of other sources is similar.

In addition to variability of a specific band, the measurement of time delays
between variations in different energy bands can also provide an important
limit to the possible values of the physical parameters in the framework of
the one-zone leptonic model.
A natural way to explain these lags is to interpret them as the cooling time
difference of the relativistic electrons
\citep[e.g.,][]{takahashi1996asca}.
In Fig.~\ref{fig7}, we plot the cooling timescales $t\mathrm{_{cool}^{obs}}$ of
different radiation processes in the observers’ frame as a function of $\gamma$ for
4FGL J1751.5+0938 (under the parameter set as shown in Table~\ref{table:parameters}).
Since the emissions of 0.1-1 GeV and 1-300 GeV
(for simplicity, we take the average values of these ranges for analysis)
usually mainly come from the EC-DT (DT is the main photon field), the corresponding
electron Lorentz factors can be estimated by using the monochromatic approximation as
\begin{equation}\label{eq26t}
\gamma\approx\left(\frac{\nu\mathrm{^{obs}}}{\nu\mathrm{^{AGN}_{DT}}}\right)^{1/2}\frac{1}{\delta},
\end{equation}
where $\nu\mathrm{^{AGN}_{DT}}$ is the peak frequency of the DT radiation.
The obtained electron Lorentz factors and the corresponding EC-DT cooling timescales
are marked with blue and red curves in Fig.~\ref{fig7}, respectively.
It can be seen that the flare in 0.1-1 GeV lags behind the flare in 1-300 GeV,
and the KN effect can be neglected, then the time lag can be given by
\begin{equation}\label{eq27t}
\Delta T\mathrm{^{obs}}=\frac{3m\mathrm{_{e}}c}{4\sigma\mathrm{_{T}}U\mathrm{_{DT}}\delta}\left(\frac{1}{\gamma_{0.1-1\mathrm{GeV}}}-\frac{1}{\gamma_{1-300\mathrm{GeV}}}\right).
\end{equation}
Substituting Eq.~(\ref{eq26t}) into Eq.~(\ref{eq27t}), $\delta$ can be derived as
\begin{equation}\label{eq28t}
\delta\approx10.6\left(\frac{1~\mathrm{day}}{\Delta T\mathrm{^{obs}}}\right)^{1/2}\left(\frac{4.5\times10^{-5}~\mathrm{erg}~\mathrm{cm}^{-3}}{U\mathrm{^{AGN}_{DT}}}\right)^{1/2}.
\end{equation}
Combining Eq.~(\ref{eq16}) and $r\mathrm{^{AGN}_{DT}}$ given by
the reverberation mapping, we have
$U\mathrm{^{AGN}_{DT}}=4.5\times10^{-5}\mathrm{~erg~cm}^{-3}/\left[1+(r\mathrm{^{AGN}}/r\mathrm{^{AGN}_{DT}})^{4}\right]$.
If assuming that the dissipation occurs within the DT
(then $U\mathrm{^{AGN}_{DT}}$ can be approximated as a constant),
and $\Delta T\mathrm{^{obs}}$ can be collected, we can obtain the value of $\delta$. If it is consistent with the lower limit of $\delta$ obtained from the internal
$\gamma\gamma$ absorption
\citep[e.g.,][]{1995MNRAS.273..583D},
we suggest that the location of the emitting region should be inside the DT,
and there should be strong EC-DT radiation in the high-energy peak,
otherwise we suggest that the emitting region should be outside the DT,
resulting in a smaller value of $U\mathrm{^{AGN}_{DT}}$,
and thus obtaining a larger value of $\delta$ to achieve consistency.
Using Eq.~(\ref{eq28t}), we have
\begin{equation}\label{eq28}
U\mathrm{_{DT}}=U\mathrm{_{DT}^{AGN}}\delta^{2}\approx5.1\times10^{-3}~\mathrm{erg~cm^{-3}}\left(\frac{1~\mathrm{day}}{\Delta T\mathrm{^{obs}}}\right).
\end{equation}
It can be seen that if $\Delta T\mathrm{^{obs}}$ is determined,
$U\mathrm{_{DT}}$ will become a constant.
Since the rough ratio of $L\mathrm{_{EC}^{obs}}/L\mathrm{_{SSC}^{obs}}/L\mathrm{_{s}^{obs}}$ can be obtained from
the quasi-simultaneous multi-wavelength SED of 4FGL J1751.5+0938 (see Fig.~\ref{fig2}),
where $L\mathrm{_{EC}^{obs}}$ and $L\mathrm{_{SSC}^{obs}}$ are the total luminosities of
the dominant EC and SSC emissions, respectively.
Combining with $U\mathrm{_{DT}}$ derived from the time delay,
we can obtain the values of $B$, $\delta$, and $r\mathrm{^{AGN}}$ successively.
More specifically, we can get the value of $B$ from
$L\mathrm{_{EC}^{obs}}/L\mathrm{_{s}^{obs}}=U\mathrm{_{DT}}/U\mathrm{_{B}}$.
Then, by combining with the causality relation, we can estimate the value of $\delta$ from
$L\mathrm{_{EC}^{obs}}/L\mathrm{_{SSC}^{obs}}=U\mathrm{_{DT}}/U\mathrm{_{syn}}$.
Finally, we can infer $r\mathrm{^{AGN}}$ from
$U\mathrm{_{DT}^{AGN}}=U\mathrm{_{DT}}/\delta^{2}$,
and then determine the origin of the high-energy peak.
On the other hand, as discussed in Section~\ref{result4.1} and shown in Fig.~\ref{fig7},
if the soft photons mainly originate from the BLR,
the cooling efficiency of the electrons producing 150 GeV emission would be greatly
reduced due to the severe KN effect, and it may result in a cooling time similar to
that of the electrons producing 500 MeV emission.
Therefore, for a specific source, if a reliable time delay between the 500 MeV and
150 GeV emissions cannot be found, we suggest that its EC process should be dominated
by EC-BLR, and then the emitting region might be located inside or outside but very
close to the BLR.
Similar to the time delay between MeV and GeV,
the time lag between flares in the infrared bands and the GeV can also help constrain
the parameters.
The difference is that the Lorentz factors of the electrons producing the infrared
emission is much lower than that producing the GeV emission,
resulting in the observed time lag being basically equal to the cooling time of the
electrons' synchrotron radiation (see Fig.~\ref{fig7}). Then, we have
$\Delta T\mathrm{^{obs}}\approx3m\mathrm{_{e}}c/\left(4\sigma\mathrm{_{T}}\gamma U\mathrm{_{B}}\delta\right)$.
It can be seen that if $\Delta T\mathrm{^{obs}}$ can be determined,
the value of $\gamma B^{2}\delta$ can be obtained.
On the other hand, we can get the value of $\gamma^{2}B\delta$ from $\nu\mathrm{^{obs}}$
corresponding to the infrared emission.
If we fix the Doppler factor to 11.7, the corresponding magnetic field
and electron Lorentz factor can be estimated
\citep[e.g.,][]{takahashi1996asca}.
Therefore, we believe that the variability timescale is very helpful for the study of
blazars.


\section{Conclusion}\label{conclusion}
We propose an analytical method to assess the necessity of external photon fields
for LBLs in the framework of one-zone scenario.
Based on obtained analytical results, we fit the quasi-simultaneous multi-wavelength
SEDs of 15 $Fermi$-4LAC LBLs with the conventional one-zone leptonic model
to investigate the physical properties of $Fermi$-4LAC LBLs.
Our main results are summarized below.
(1) We find that most of the LBLs’ SEDs cannot be fitted by the one-zone SSC model,
which implies that the introduction of external photon fields is almost inevitable
for the explanation of the high-energy peak. 
Therefore, our results indicate that LBLs are masquerading BL Lacs, confirming the
difference in physical properties between LBLs and IBLs/HBLs
\citep[e.g.,][]{2014MNRAS.439.2933Y, 2012ApJ...761..125F}.
(2) We suggest that the $\gamma$-ray emitting regions of LBLs are located
outside the BLR and within the DT.
(3) By extending the analytical method to all types of LSPs in $Fermi$-4LAC
\citep[using archival data;][]{2022ApJS..262...18Y, 2023SCPMA..6649511Y},
we find that the high-energy peaks of some FSRQs and BCUs can be explained by the
SSC emission, implying that the importance of external photons could be minor.
(4) Our results suggest that the variability timescale may be a useful indicator to
the origin of the high-energy peak, and the high-energy peak is only possible to
originate from the SSC emission in the case of relatively long variability timescale.

\section*{Acknowledgements}
This work is partially supported by the National Key Research and Development Program of China (grant numbers 2022SKA0130100 and 2022SKA0130102). R.X. acknowledges the support by the NSFC under Grant No. 12203043. F.K.P acknowledges support by the National Natural Science Foundation of China (Grants No. 12003002), the University Annual Scientific Research Plan of Anhui Province 2023 (2023AH050146), the Excellent Teacher Training Program of Anhui Province (2023), and the Doctoral Starting up Foundation of Anhui Normal University 2020 (903/752022). Part of this work is based on archival data, software, or online services provided by the SPACE SCIENCE DATA CENTER (SSDC).

\section*{Data Availability}
The data underlying this article will be shared on reasonable request to the corresponding author.



\bibliographystyle{mnras}
\bibliography{example} 








\bsp	
\label{lastpage}
\end{document}